\documentclass[prd,preprint,tightenlines,nofootinbib,eqsecnum,showpacs]{revtex4-1}

\usepackage{amsmath}
\usepackage{amsfonts}
\usepackage{amssymb}
\usepackage{mathrsfs}
\usepackage{graphicx}
\usepackage{hyperref}
\usepackage{color}
\usepackage{array}
\usepackage{mathtools}
\usepackage{booktabs}

\newcolumntype{C}{>{$}c<{$}}
\AtBeginDocument{
\heavyrulewidth=.08em
\lightrulewidth=.05em
\cmidrulewidth=.03em
\belowrulesep=.65ex
\belowbottomsep=0pt
\aboverulesep=.4ex
\abovetopsep=0pt
\cmidrulesep=\doublerulesep
\cmidrulekern=.5em
\defaultaddspace=.5em
}

\definecolor{CiteColor}{rgb}{0,0,0.35}
\definecolor{URLColor}{rgb}{0,0,0.35}
\hypersetup{colorlinks,citecolor=CiteColor,urlcolor=URLColor}

\newcommand{\beq}{\begin{equation}}
\newcommand{\eeq}{\end{equation}}
\newcommand{\ud}{\mathrm{d}}
\newcommand{\calD}{\mathcal{D}}
\newcommand{\calH}{\mathcal{H}}
\newcommand{\calO}{\mathcal{O}}
\newcommand{\av}[1]{\langle #1 \rangle}

\allowdisplaybreaks

\begin{document}

\title{Horizon Surface Gravity in Corotating Black Hole Binaries}

\author{Alexandre \textsc{Le Tiec}}
\affiliation{LUTH, Observatoire de Paris, PSL Research University, CNRS,
Universit\'e Paris Diderot, Sorbonne Paris Cit\'e, 92190 Meudon, France}

\author{Philippe \textsc{Grandcl\'ement}}
\affiliation{LUTH, Observatoire de Paris, PSL Research University, CNRS,
Universit\'e Paris Diderot, Sorbonne Paris Cit\'e, 92190 Meudon, France}

\date{\today}

\begin{abstract}
For binary systems of corotating black holes, the zeroth law of black hole mechanics states that the surface gravity is constant over each component of the horizon. Using the approximation of a conformally flat spatial metric, we compute sequences of quasi-equilibrium initial data for corotating black hole binaries with irreducible mass ratios in the range $10:1-1:1$. For orbits outside the innermost stable one, the surface gravity is found to be constant on each component of the apparent horizon at the sub-percent level. We compare those numerical results to the analytical predictions from post-Newtonian theory at the fourth (4PN) order and from black hole perturbation theory to linear order in the mass ratio. We find a remarkably good agreement for all mass ratios considered, even in the strong-field regime. In particular, our findings confirm that the domain of validity of black hole perturbative calculations appears to extend well beyond the extreme mass-ratio limit. 
\end{abstract}


\maketitle

\section{Introduction and summary}

The first detections of gravitational waves from coalescing black hole binaries has herald a new era in astronomy \cite{Ab.al2.16,Ab.al3.16,Ab.al.17,Ab.al2.17}. By the end of this decade, an increasingly sensitive network of terrestrial laser interferometers (LIGO, Virgo, KAGRA) will detect up to several hundreds of such events per year \cite{Ab.al.18}. Those gravitational-wave observations will not only tell us about the underlying population of binary black holes, but also allow testing the general theory of relativity (GR), and improve our understanding of the dynamics of compact binary systems in a highly dynamical, strong-field regime \cite{Ab.al4.16}.

In this paper, we focus on \textit{corotating} black hole binaries, which correspond to the particular case where the black holes are rotating synchronously with the orbital motion. Heuristically, these can be viewed as stationary in a corotating frame. A great deal of attention has already been devoted to such systems \cite{Go.al.02,Gr.al.02,Fr.al.02,Co2.02,Bl.02,Da.al.02,CoPf.04,Kl.04,Ca.al2.06}, because they represent the only configuration of two black holes that can possess a true helical Killing field, and for which the rotation state of each black hole is fully constrained. Interestingly, the exact helical symmetry allows defining the horizon surface gravity of each black hole in the standard way \cite{Poi}. Moreover, according to the zeroth law of binary black hole mechanics \cite{Fr.al.02}, the surface gravity must be constant over each component of the horizon.

While sequences of quasi-equilibrium initial data for corotating black hole binaries have been constructed long ago, the main interest at the time was to extract the binding energy of the system, as a function of the circular-orbit frequency, in order to locate the innermost circular orbit. Here we revisit this problem, but focus instead on the horizon surface gravity of each black hole, which, rather surprisingly, has never been investigated in this context. Using the standard approximation of a conformally flat spatial metric, we compute sequences of quasi-equilibrium initial data for corotating black hole binaries with irreducible mass ratios $10:1$, $2:1$ and $1:1$, and extract the surface gravity of each black hole.

On the other hand, if the orbital separation is large, then such systems can be investigated in the context of the post-Newtonian (PN) approximation to GR. Each black hole can then be modelled as a spinning point particle. Various arguments indicate that the point-particle analogue of the horizon surface gravity of a black hole is the ``redshift variable,'' a conserved quantity associated with the helical symmetry \cite{Le.al.12,Bl.al.13,Po3.15,Zi.al.16}. Building upon the recent derivation of the 4PN equations of motion for binary systems of nonspinning point particles \cite{Da.al.14,Be.al.16,Ma.al.18}, we add the relevant spin-orbit contributions and compute the redshifts in corotating point-particle binaries up to 4PN order,\footnote{As usual we refer to $n$PN as the order corresponding to terms $\calO(c^{-2n})$ in the equations of motion.} for comparison to the numerical results.

Corotating black hole binaries have also been studied in the limit of extreme mass ratios, in the context of linear black hole perturbation theory. The authors of Ref.~\cite{GrLe.13} showed that, for a Kerr black hole perturbed by a small corotating moon, the perturbed event horizon is a Killing horizon of the helical Killing field. The surface gravity is constant over the horizon, and the perturbation in surface gravity was computed analytically, together with the perturbation in the horizon angular velocity \cite{GrLe.13}. Combining those results, we compute in closed form the surface gravity of the larger body, to linear order in the mass ratio. We also compute the redshift of the corotating point particle, i.e., the analogue of the horizon surface gravity of the smaller body, still to linear order in the mass ratio.

A growing body of work \cite{FiDe.84,An.al.95,Fa.al.04,Sp.al2.11,Le.al.11,Le.al2.12,Le.al.13,Na.13,vdM.17} suggests that the domain of validity of black hole perturbative calculations can be extended well beyond the extreme mass-ratio limit, by means of a simple but physically-motived rescaling of the component masses. Here, we perform such a rescaling and show that the rescaled perturbative predictions are in remarkable agreement with the numerical relativity (NR) results obtained by solving (part of) the Einstein field equations. As pointed out in Ref.~\cite{Le2.14}, this opens the exciting prospect of using black hole perturbation theory to model the gravitational-wave emission from intermediate mass-ratio inspirals or even compact binaries with comparable masses.

The remainder of this paper is organized as follows. In Sec.~\ref{sec:laws} we give a summary of the known results on the laws of mechanics for single and binary black hole systems. Section \ref{sec:NR} details our computations of quasi-equilibrium initial data for corotating black hole binaries with mass ratios in the range $10:1-1:1$, from which we extract the surface gravity. We then compute the (analogue of the) surface gravity for corotating point particles in Sec.~\ref{sec:PN}, in the context of the PN approximation, and for a Kerr black hole perturbed by a corotating moon in Sec.~\ref{sec:BHPT}, using linear perturbation theory. Finally, we compare those numerical and analytical results in Sec.~\ref{sec:compare}.

Throughout this paper our conventions are those of \cite{Wal}. The metric signature is $(-+++)$ and we use ``geometrized units'' where $G = c = 1$. Greek indices $\alpha,\beta,\dots$ are abstract, while Latin indices $i,j,\dots$ are used for spatial coordinate components in a given coordinate system.

\section{Mechanics of single and binary black holes}
\label{sec:laws}

In this section, we provide a summary of known results on the laws of mechanics for single (isolated) and binary black hole systems. For the convenience of the reader, we collect in one place many results that were previously established in several papers. We discuss the various zeroth and first laws of mechanics that have been derived for isolated, stationary black holes in Sec.~\ref{subsec:1BH}, for binary systems of corotating black holes in Sec.~\ref{subsec:2BH}, for a spinning black hole perturbed by a corotating moon in Sec.~\ref{subsec:BHPT}, for two spinning point particles in Sec.~\ref{subsec:2PP}, and for two corotating point particles in Sec.~\ref{subsec:2PPcor}. An analogy between the horizon surface gravity of a black hole and the redshift of a point particle is discussed in Sec.~\ref{subsec:z-kappa}.

\subsection{Single stationary black hole}\label{subsec:1BH}

First, we consider an isolated, stationary black hole of mass $M$, angular momentum $S$, and horizon surface area $A$. Hereafter, we will frequently use the irreducible mass $\mu = \sqrt{A/ (16 \pi)}$ instead of the horizon surface area $A$ in various formulas.

\subsubsection{Zeroth law}\label{subsubsec:0thlaw}

The event horizon of a stationary black hole is a Killing horizon, i.e., a null hypersurface $\calH$ whose geodesic generators are tangent to a Killing vector field $k^\alpha$. For any Killing horizon, the surface gravity $\kappa$ is the scalar field defined on $\calH$ by
\beq\label{kappa2}
	\kappa^2 \equiv - \frac{1}{2} \nabla^\alpha k^\beta \, \nabla_\alpha k_\beta \big|_\calH \, .
\eeq
Assuming the dominant energy condition, Bardeen, Carter and Hawking \cite{Ba.al.73} established that the scalar field \eqref{kappa2} must be constant over $\calH$. The zeroth law of black hole mechanics then states that the surface gravity $\kappa$ of a stationary black hole is constant over its event horizon $\calH$. The constant $\kappa$ is also known as the non-affinity coefficient, because it is directly related to the failure of the null geodesic generators to be affinely parameterized:
\beq
	k^\alpha \nabla_\alpha k^\beta \vert_\calH = \kappa \, k^\beta \vert_\calH \, .
\eeq

Specializing to a Kerr black hole of mass $M$ and angular momentum (or spin) $S$, we denote by $t^\alpha$ the timelike Killing field normalized to $-1$ at infinity, and by $\phi^\alpha$ the axial Killing field with integral curves of parameter length $2\pi$. The horizon-generating Killing field $k^\alpha$ must be a linear combination of $t^\alpha$ and $\phi^\alpha$, the only two generators of isometries of the Kerr metric. Choosing the normalization of $k^\alpha$ such that $k^\alpha t_\alpha = -1$ at spatial infinity, we thus have
\beq
	k^\alpha = t^\alpha + \omega \, \phi^\alpha \, ,
\eeq
where the constant $\omega$ can be interpreted as the angular velocity of the horizon's generators, namely the apparent rotation rate of the black hole, as measured by a static, inertial observer at infinity. With this choice of normalization, the condition $k^\alpha k_\alpha \vert_\calH = 0$ and the definition \eqref{kappa2} yield the following expressions for the constant surface gravity and angular velocity of a Kerr black hole \cite{Poi}:
\begin{subequations}\label{omega_kappa_Kerr}
	\begin{align}
		\kappa &= \frac{\sqrt{1-\chi^2}}{2M \bigl( 1 + \sqrt{1-\chi^2} \bigr)} \, , \label{kappa_Kerr} \\
		\omega &= \frac{\chi}{2M \bigl( 1 + \sqrt{1-\chi^2} \bigr)} \, , \label{omega_Kerr}
	\end{align}
\end{subequations}
where $\chi \equiv S/M^2$ is the dimensionless Kerr parameter, such that $0 \leqslant \chi < 1$. For a nonspinning black hole, $\omega = 0$ and $\kappa = 1/(4M)$, as expected from a (naive) Newtonian calculation of the surface gravity of a spherically symmetric body of mass $M$ and typical radius $2M$.

\subsubsection{First law}

On the other hand, the celebrated first law of black hole mechanics of Bardeen \textit{et al.} \cite{Ba.al.73} relates the changes in the mass, spin and surface area of nearby stationary and axisymmetric black holes solutions according to\footnote{Recalling the relation $A = 16\pi \mu^2$ between the surface area $A$ and the irreducible mass $\mu$, the last term in the right-hand side of Eq.~\eqref{1st_law} could also be written as $4 \kappa \mu \, \delta \mu$.}
\beq\label{1st_law}
	\delta M = \omega \, \delta S + \frac{\kappa}{8\pi} \, \delta A \, .
\eeq
Iyer and Wald \cite{IyWa.94} gave a strengthened form of this result, proving that it holds for arbitrary nonsingular, vacuum perturbations of a stationary black hole that are asymptotically flat at spatial infinity and regular on the event horizon $\calH$. Moreover, a ``physical process version'' of the first law \eqref{1st_law} was derived \cite{GaWa.01}, and extensions to electrovac and non-vacuum spacetimes are known as well \cite{Ca.10}. By noticing that the mass $M$ is a homogeneous function of degree one in the variables $S^{1/2}$ and $A^{1/2}$, it can be shown that there exists a ``first integral'' associated with the variational formula \eqref{1st_law}, namely \cite{Sm.73}
\beq\label{Smarr}
	M = 2 \omega S + \frac{\kappa A}{4\pi} \, .
\eeq

Hereafter, it will prove convenient to have the expressions for the mass $M$ and spin $S$ as functions of the irreducible mass $\mu$ and the horizon angular velocity $\omega$. Combining the first law \eqref{1st_law} with the Christodoulou mass formula $M^2 \!=\! \mu^2 + S^2 / (4\mu^2)$ for Kerr black holes \cite{Ch.70}, it can easily be shown that
\begin{subequations}\label{M-S}
	\begin{align}
		M &= \frac{\mu}{\sqrt{1 - 4(\mu \omega)^2}} \, , \\
		S &= \frac{4\mu^3 \omega}{\sqrt{1 - 4(\mu \omega)^2}} \, . \label{S}
	\end{align}
\end{subequations}
In the slow rotation limit $\mu \omega \ll 1$, one finds the expression $S \simeq 4\mu^3 \omega$ for the black hole spin, as expected for a body of mass $M \simeq \mu$ with typical radius $2M$ and rotational velocity $2M \omega$. Then, substituting for Eqs.~\eqref{M-S} into the formula \eqref{kappa_Kerr}, we obtain the following expression for the normalized surface gravity $4\mu\kappa$ of a Kerr black hole as a function of $\mu$ and $\omega$:
\beq\label{4mukappa_Kerr}
	4\mu\kappa = \frac{1 - 8(\mu \omega)^2}{\sqrt{1 - 4(\mu \omega)^2}} \, .
\eeq

\subsection{Two corotating black holes}\label{subsec:2BH}

In a seminal work, Friedman, Ury\=u and Shibata \cite{Fr.al.02} generalized the zeroth and first laws discussed above to spacetimes containing multiple black holes and/or a generic distribution of perfect fluid matter sources with compact support, having in mind the application of their general results to binary systems of black holes and/or neutron stars.\footnote{See Ref.~\cite{Ur.al.10} for the generalization of their work to the magnetohydrodynamic case.}

\subsubsection{Zeroth law}

We consider vacuum, binary black hole spacetimes endowed with a global \textit{helical} Killing vector $k^\alpha$.\! Following Refs.\!~\cite{Go.al.02,Fr.al.02}, a spacetime is said to have a helical Killing symmetry if the generator of the isometry can be written in the form
\beq\label{helical_k}
	k^\alpha = t^\alpha + \Omega \, \phi^\alpha \, ,
\eeq
where $\Omega > 0$ is a constant, $t^\alpha$ is timelike and $\phi^\alpha$ is spacelike with circular orbits of parameter length $2\pi$. In general, neither $t^\alpha$ nor $\phi^\alpha$ is a Killing vector, but the combination \eqref{helical_k} is a Killing vector for a specific value of the constant $\Omega$, which can be interpreted as the angular frequency of the binary black hole system.

Physically, the only allowed configurations correspond to a state of \textit{corotation}, for which the null geodesic generators of each component of the horizon are tangent to the Killing vector field \eqref{helical_k}. Indeed, if corotation were not realized, the resulting non-vanishing shear would result in the growth of the horizon's surface areas, in contradiction with the hypothesis of helical symmetry. For such corotating binaries, each component $\calH_a$ of the event horizon is a Killing horizon with horizon-generating Killing field $k^\alpha \vert_{\calH_a}$. The authors of Ref.~\cite{Fr.al.02} could then show that the surface gravity $\kappa_a$, as defined by \eqref{kappa2}, is indeed constant on $\calH_a$:
\beq\label{0th_law}
	\kappa_a^2 \equiv - \frac{1}{2} \nabla^\alpha k^\beta \, \nabla_\alpha k_\beta \big|_{\calH_a} = \text{const.}
\eeq

However, in GR it is known that helically symmetric spacetimes cannot be asymptotically flat \cite{GiSt.84,De.89,Kl.04}. This fact can easily be understood from a heuristic point of view: in order to maintain the binary on a fixed circular orbit, the energy radiated in gravitational waves needs to be compensated by an equal amount of incoming radiation. Far away from the source, the resulting system of standing waves ends up dominating the energy content of the spacetime, such that the falloff conditions necessary to ensure asymptotic flatness cannot be satisfied. This lack of asymptotic flatness implies that there is no natural normalization of the helical Killing field $k^\alpha$, and hence the numerical value of the surface gravity $\kappa_a$ of each black hole is entirely free.

Nevertheless, asymptotic flatness can be recovered if, loosely speaking, the gravitational radiation can be ``turned off.'' This can be achieved, in particular, using the Isenberg, Wilson and Mathews approximation to GR, also known as the conformal flatness condition (CFC) approximation \cite{IsNe.80,WiMa.89,Is.08}. In Sec.~\ref{sec:NR}, we will rely on this approximation to compute sequences of quasi-equilibrium initial data for corotating black hole binaries, and to properly normalize the surface gravity of each black hole.

\subsubsection{First law}

Consider a vacuum spacetime containing multiple black holes, and endowed with a \textit{global} Killing vector field $k^\alpha$. The Noether current associated with $k^\alpha$ assigns to each spacetime a conserved charge $Q$. The main result established in Ref.~\cite{Fr.al.02} relates the variation $\delta Q$ of the conserved charge to the variations $\delta A_a$ of the black hole's horizon surface areas, namely
\beq\label{1st_law_gen}
	\delta Q = \sum_a \frac{\kappa_a}{8\pi} \, \delta A_a \, ,
\eeq
where $\kappa_a$ is the constant horizon surface gravity of the $a$\textsuperscript{th} black hole. When matter sources are present, additional terms appear in the right-hand side of this formula. We are interested in applying the general result \eqref{1st_law_gen} to the particular case of a \textit{binary} system of corotating black holes moving along circular orbits.

For such spacetimes, the geometry is invariant along the integral curve of a helical Killing vector field \eqref{helical_k}. In general, neither $t^\alpha$ nor $\phi^\alpha$ are Killing vectors. However, for asymptotically flat spacetimes $t^\alpha$ and $\phi^\alpha$ are asymptotically Killing, and the variation of the conserved charge $Q$ takes the form $\delta Q = \delta M_\text{ADM} - \Omega \, \delta J$ \cite{Fr.al.02}, where $M_\text{ADM}$ is the Arnowitt-Deser-Misner (ADM) mass and $J$ is the total angular momentum, both defined as surface integrals at spatial infinity. Then, the general formula \eqref{1st_law_gen} reduces to the first law of binary black hole mechanics:
\beq\label{1st_law_2BH}
	\delta M_\text{ADM} = \Omega \, \delta J + \sum_a \frac{\kappa_a}{8\pi} \, \delta A_a \, .
\eeq
Notice that, in the binary black hole case, the horizon angular velocity $\omega$ of a single rotating black hole is replaced by the orbital frequency $\Omega$ of the binary system.

Just like the first law \eqref{1st_law} for a single black hole admits a ``first integral,'' namely Smarr's formula \eqref{Smarr}, in the binary black hole case there is also a first integral associated with the variational formula \eqref{1st_law_2BH}, namely \cite{Le.al.12}
\beq\label{1st_int_2BH}
	M_\text{ADM} = 2 \Omega J + \sum_a \frac{\kappa_a A_a}{4\pi} \, . 
\eeq
This algebraic relationship will play a key role in Sec.~\ref{sec:NR}, by allowing us to test the numerical accuracy of our sequences of quasi-equilibrium initial data for corotating binary black holes.

\subsection{Black hole and corotating moon}\label{subsec:BHPT}

Gralla and Le Tiec \cite{GrLe.13} have shown how the zeroth and first laws of black hole mechanics can be extended to the case where a Kerr black hole is perturbed by a small orbiting compact object (a ``moon'') on the \emph{corotating} orbit, the unique circular equatorial orbit whose orbital frequency matches the angular velocity of the Kerr horizon. The natural framework for such a problem is that of black hole perturbation theory. Therefore, in this section we consider a Kerr black hole of mass $M$ and spin $S$, perturbed by a corotating point particle of mass $m \ll M$, and we work to linear order in the small mass ratio $q \equiv m/M$.

\subsubsection{Zeroth law}

The metric perturbation generated by the corotating moon is helically symmetric, asymptotically flat at future null infinity,\footnote{By working to linear order in the mass ratio, the metric perturbation is (in a sense) taken to be ``infinitesimally small,'' such than an helically symmetric spacetime can be asymptotically flat as well \cite{Ke.al2.10}.} and regular on the black hole's perturbed event horizon $\calH$. Combining those properties, it can be shown that the expansion and shear of $\calH$ must vanish, thus ensuring (through rigidity arguments) that the \textit{perturbed} event horizon is a Killing horizon \cite{GrLe.13}. The horizon-generating Killing field, say $k^\alpha$, must then be a linear combination of (i) the helical Killing field $\ell^\alpha$ of the perturbed spacetime, simply given by $\ell^\alpha = t^\alpha + \bar{\omega} \, \phi^\alpha$ in adapted coordinates, where $\bar{\omega}$ is the angular velocity of the background Kerr black hole [recall Eq.~\eqref{omega_Kerr}], and (ii) perturbations of the background Killing fields inherited from the isometries of the Kerr geometry. Since the perturbed spacetime is asymptotically Minkowskian at future null infinity, the horizon-generating Killing field $k^\alpha$ may be normalized there according to
\beq\label{k}
	k^\alpha = t^\alpha + \bigl( \bar{\omega} + \calD \omega \bigr) \, \phi^\alpha \, ,
\eeq
where the constant $\calD \omega = \mathcal{O}(q)$ is the perturbation in angular velocity induced by the moon. By analogy with the binary black hole case, the constant $\Omega \equiv \bar{\omega} + \calD \omega$ will be referred to as the circular-orbit frequency of the binary system.

Since $\calH$ is a Killing horizon, the surface gravity $\kappa$ of the perturbed black hole, as defined by \eqref{kappa2}, is constant over $\calH$. It can be written in the form $\kappa = \bar{\kappa} + \calD \kappa$, where $\bar{\kappa}$ is the surface gravity of the background Kerr black hole [recall Eq.~\eqref{kappa_Kerr}], and $\calD \kappa = \mathcal{O}(q)$ is the constant perturbation induced by the corotating moon. Remarkably, simple closed-form expressions for the perturbations $\calD \kappa$ and $\calD \omega$ in surface gravity and angular velocity can be established, namely \cite{GrLe.13}
\begin{subequations}\label{Dkappa_Domega}
	\begin{align}
		\calD \kappa &= z \, \bar{\kappa} \, \frac{\partial H}{\partial M} \, , \label{Dkappa} \\
		\calD \omega &= z \, \bar{\omega} \, \frac{\partial H}{\partial M} + z \, \frac{\partial H}{\partial S} \, , \label{Domega}
	\end{align}
\end{subequations}
where $z \equiv - k^\alpha u_\alpha$ is the ``redshift'' of the particle, the conserved orbital quantity associated with the helical Killing symmetry of the perturbed spacetime, with $u^\alpha$ the particle's four-velocity. The formulas \eqref{Dkappa_Domega} involve the canonical Hamiltonian $H$ that controls the timelike geodesic motion of a \textit{test} particle of mass $m$ in the Kerr geometry. 

An explicit calculation reveals that the relative changes in horizon surface gravity and angular velocity induced by the orbiting moon are given by the expressions \cite{Le.12,GrLe.13}
\begin{subequations}\label{Dkappa-Domega}
	\begin{align}
		\frac{\calD \kappa}{\bar{\kappa}} &= - q \,\, \frac{v^2}{1 + \chi v^3} \, \frac{1 + 2 \chi v^3 - \chi^2 v^4}{\sqrt{1 - 3 v^2 + 2 \chi v^3}} \, , \label{Dkappa2} \\
		\frac{\calD \omega}{\bar{\omega}} &= q \,\, \frac{v^2}{1 + \chi v^3} \, \frac{2s v^3 - \chi (1 + s v^4)}{\chi \sqrt{1 - 3 v^2 + 2 \chi v^3}} \, , \label{Domega2}
	\end{align}
\end{subequations}
where we introduced the shorthands $v^3 \!\equiv\! M \Omega / (1 - \chi M \Omega)$ and $s \!\equiv\! 2 + 2 \sqrt{1-\chi^2} - \chi^2$. At this order of approximation, we may substitute $\Omega$ by $\bar{\omega}$ and use the relation \eqref{omega_Kerr} between $\bar{\omega}$ and $\chi$, so that the relative changes \eqref{Dkappa-Domega} are functions of $\chi$ (or $\bar{\omega}$) only. Notice that the change in surface gravity induced by the tidal field of the moon is negative, while the change in horizon angular velocity is positive. Hence, from the point of view of a distant inertial observer, the apparent rotation rate of the black hole is increased by the corotating moon.

\subsubsection{First law}

Adapting Iyer's and Wald's formalism \cite{IyWa.94} to nonvacuum perturbations of a nonstationary black hole spacetime that are asymptotically flat at future null infinity, the authors of \cite{GrLe.13} could generalize the first law \eqref{1st_law} to binary systems of spinning black holes with corotating moons. Their first law relates variations in the Bondi mass $M_\text{B}$ and angular momentum $J_\text{B}$ of the perturbed spacetime, as measured at future null infinity, to variations in the perturbed black hole's surface area $A$ and the moon's mass $m$, and reads\footnote{The symbol $\delta$, used to denote a variation within the three-parameter family of black hole with corotating moon spacetimes, should not be confused with the symbol $\calD$, used to denote a small change induced by the nonvanishing mass $m$ of the orbiting moon.}
\beq\label{1st_law_BH-PP}
	\delta M_\text{B} = \Omega \, \delta J_\text{B} + \frac{\kappa}{8\pi} \, \delta A + z \, \delta m \, ,
\eeq
where we recall that $z$ is the particle's ``redshift.'' Here, $\kappa = \bar{\kappa} + \calD \kappa$ is the constant horizon surface gravity of the perturbed black hole, while $\Omega = \bar{\omega} + \calD \omega$ is the circular-orbit frequency of the binary [recall Eq.~\eqref{k}]. The derivation of the first law \eqref{1st_law_BH-PP} is analogous to that outlined in Sec.~\ref{subsec:2BH}. In particular, the combination $\delta M_\text{B} - \Omega \, \delta J_\text{B}$ appears when evaluating the variation of the conserved Noether charge associated with the helical Killing vector field $k^\alpha = t^\alpha + \Omega \, \phi^\alpha$, except that the surface integral is performed on a two-sphere at future null infinity, and not at spatial infinity.

Moreover, just like the first laws \eqref{1st_law} and \eqref{1st_law_2BH} for isolated black holes and corotating black hole binaries admit first integral relations, there is a first integral associated with the first law \eqref{1st_law_BH-PP} for a black hole with a corotating moon, namely \cite{GrLe.13}
\beq\label{1st_integral_BH-PP}
	M_\text{B} = 2 \Omega J_\text{B} + \frac{\kappa A}{4\pi} + z m \, .
\eeq
This can be established either by applying Stokes' theorem to the Noether charge associated with $k^\alpha$, or by combining Euler's theorem for homogeneous functions with Eq.~\eqref{1st_law_BH-PP}. Note that the first law \eqref{1st_law_BH-PP} and the first integral \eqref{1st_integral_BH-PP} were established up to relative $\calO(q)$.

\subsection{Two spinning point particles}\label{subsec:2PP}

Next, we consider binary systems of compact objects, modelled as (spinning) point particles with constant masses $m_a$. Although the concept of a point particle does not make sense in exact GR \cite{GeTr.87}---the closest thing to a point particle in GR is a black hole---, in the context of \textit{approximation schemes} such as PN theory \cite{Bl.14}, Einstein's equation can be coupled to a distributional matter source in a meaningful manner, provided that a regularization scheme (e.g. dimensional regularization) is used to handle the divergent self-field of each particle.

Since a point particle has no ``spatial'' extension, and in particular no horizon, a putative zeroth law of mechanics is meaningless for binary systems of point particles. Still, a first law can be derived by assuming the existence of a global helical Killing field $k^\alpha$, and by taking a formal point-particle limit in the generalized first law of \cite{Fr.al.02}, in the case of two self-gravitating balls of perfect fluid. The resulting first law of binary point-particle mechanics reads \cite{Le.al.12}
\beq\label{1st_law_2PP}
	\delta M_\text{ADM} = \Omega \, \delta J + \sum_a z_a \, \delta m_a \, ,
\eeq
where $z_a \equiv -u_a^\alpha k_\alpha$ is the ``redshift'' of particle $a$, a conserved orbital quantity associated with the helical Killing symmetry, with $u_a^\alpha$ the four-velocity of particle $a$. Choosing coordinates adapted to that isometry, such that $k^\alpha = {(\partial_t)}^\alpha + \Omega \, {(\partial_\phi)}^\alpha$ holds everywhere, and not merely in a neighbourhood of infinity, the redshift of particle $a$ is simply $z_a = \ud \tau_a / \ud t$, the ratio of the proper times elapsed along the worldline of the particle and along that of a distant static observer. As expected, the first integral associated with the variational formula \eqref{1st_law_2PP} is
\beq
	M_\text{ADM} = 2\Omega J + \sum_a z_a m_a \, .
\eeq

The first law \eqref{1st_law_2PP} was later recovered, and extended to \textit{spinning} point particles, using the canonical ADM Hamiltonian framework \cite{Bl.al.13}. Assuming that the conservative dynamics of a binary system of spinning point particles with masses $m_a$ and canonical spins $\mathbf{S}_a$ derives from an autonomous Fokker-type two-body Hamiltonian $H_\text{F}(\mathbf{x}_a,\mathbf{p}_a;m_a,\mathbf{S}_a)$, with $\mathbf{x}_a$ and $\mathbf{p}_a$ the canonical positions and momenta, the first law for circular orbits and spins aligned or anti-aligned with the orbital angular momentum reads\footnote{Strictly speaking, the mass appearing in the left-hand side of Eq.~\eqref{1st_law_2PPspin} is the on-shell value of the local, autonomous Fokker Hamiltonian $H_\text{F}(\mathbf{x}_a,\mathbf{p}_a;m_a,\mathbf{S}_a)$.}
\beq\label{1st_law_2PPspin}
	\delta M_\text{ADM} = \Omega \, \delta J + \sum_a \left[ z_a \, \delta m_a + (\Omega_a - \Omega) \, \delta S_a \right] .
\eeq
Interestingly, while the precession frequencies $\Omega_a$ of the spins are given by the partial derivatives of the two-body Hamiltonian with respect to the particle's spins \cite{Da.al.08}, it was shown in Ref.~\cite{Bl.al.13} that (at least to linear order in the spins) the redshifts $z_a$ are given by the partial derivatives of the Hamiltonian with respect to the particle's masses:
\beq\label{tic-tac}
	\Omega_a = \frac{\partial H_\text{F}}{\partial S_a} \, , \qquad z_a = \frac{\partial H_\text{F}}{\partial m_a} \, .
\eeq
Using the canonical ADM framework, the first law \eqref{1st_law_2PP} for circular orbits was also extended to generic bound (i.e. eccentric) orbits in Refs.~\cite{Le.15,BlLe.17}. These first laws for binary systems of (spinning) point masses, as well as the associated first integrals, have been explictly checked to hold true up to 3PN order included, as well as the logarithmic contributions at the 4PN and 5PN orders.

\subsection{Two corotating point particles}\label{subsec:2PPcor}

Motivated by the first law \eqref{1st_law_2BH} for binary systems of corotating black holes, the authors of Ref.~\cite{Bl.al.13} suggested modelling each spinning point particle as an equilibrium black hole in a tidal environment, endowed with an ``irreducible mass'' $\mu_a$ and a ``proper rotation frequency'' $\omega_a$. Then, by analogy with the first law \eqref{1st_law} for a single, isolated black hole, each spinning point particle with mass $m_a$ and spin $S_a$ was assumed to obey the variational formula
\beq\label{ploup}
	\delta m_a = c_a \, \delta \mu_a + \omega_a \, \delta S_a \, .
\eeq
To obtain explicit expressions for the coefficients $c_a = (\partial m_a / \partial \mu_a)_{S_a}$ and $\omega_a = (\partial m_a / \partial S_a)_{\mu_a}$, it was further assumed that each spinning point particle obeys a Christodoulou mass formula of the type $m_a^2 = \mu_a^2 + S_a^2 / (4\mu_a^2)$. Then, for each particle the ``response coefficient'' $c_a$, which is the point-particle analogue of the normalized surface gravity $4\mu\kappa$ of an isolated black hole, can be expressed as a function of e.g. $\mu_a$ and $\omega_a$, and similarly for the mass $m_a$ and spin $S_a$. From Eqs.~\eqref{M-S} and \eqref{4mukappa_Kerr}, we immediately find
\begin{subequations}\label{ma-Sa-ca}
	\begin{align}
		m_a &= \frac{\mu_a}{\sqrt{1 - 4(\mu_a \omega_a)^2}} \, , \label{ma} \\
		S_a &= \frac{4\mu_a^3 \omega_a}{\sqrt{1 - 4(\mu_a \omega_a)^2}} \, , \label{Sa} \\
		c_a &= \frac{1 - 8(\mu_a \omega_a)^2}{\sqrt{1 - 4(\mu_a \omega_a)^2}} \equiv 4 \mu_a \bar{\kappa}_a \, . \label{ca}
	\end{align}
\end{subequations}

Substituting for Eq.~\eqref{ploup} into the first law \eqref{1st_law_2PPspin} for binary systems of spinning point particles yields a formula that involves the variations $\delta \mu_a$ and $\delta S_a $ of the irreducible masses and the spins, namely $\delta M_\text{ADM} = \Omega \, \delta J + \sum_a \left[ c_a z_a \, \delta \mu_a + (\Omega_a - \Omega + z_a \omega_a) \, \delta S_a \right]$. By analogy with the first law \eqref{1st_law_2BH} for binary systems of corotating black holes, the proper rotation state of each spinning particle is constrained to a corotation state if, and only if, the coefficients in front of the variations $\delta S_a$ vanish, yielding the corotation conditions (for $a=1,2$)
\beq\label{cond_cor}
	z_a \, \omega_a = \Omega - \Omega_a \, .
\eeq
Physically, this means that the redshifted proper rotation frequency of each spinning point particle, $z_a \omega_a$, must be equal to the circular orbit frequency $\Omega$, as seen in a frame rotating at the angular rate $\Omega_a$ with respect to an inertial frame of reference.

When the corotating condition \eqref{cond_cor} is imposed, the first law \eqref{1st_law_2PPspin} for binary systems of spinning point particles reduces to
\beq\label{1st_law_2PP_cor}
	\delta M_\text{ADM} = \Omega \, \delta J + \sum_a c_a z_a \, \delta \mu_a \, .
\eeq
Because the first law \eqref{1st_law_2BH} for corotating black hole binaries can be written in the equivalent form $\delta M_\text{ADM} = \Omega \, \delta J + \sum_a 4 \mu_a \kappa_a \, \delta \mu_a$, it appears that the renormalized redshifts $c_a z_a$ in corotating point particle binaries are analogous to the normalized surface gravities $4\mu_a \kappa_a$ in corotating black hole binaries. This will be further discussed in the next Sec.~\ref{subsec:z-kappa}. To conclude this section, we point out that the first integral relation associated with the variational first law \eqref{1st_law_2PP_cor} reads, as expected,
\beq\label{1st_int_2PP_cor}
	M_\text{ADM} = 2 \Omega J + \sum_a c_a z_a \mu_a \, .
\eeq

\subsection{Surface gravity and redshift}\label{subsec:z-kappa}

We can now compare the first law \eqref{1st_law_2BH} for corotating black hole binaries to the first law \eqref{1st_law_2PP_cor} for corotating point particles binaries, as well as the first integrals \eqref{1st_int_2BH} and \eqref{1st_int_2PP_cor}. Clearly, if the irreducible masses $\mu_a$ are identified for each body, then the surface gravity $\kappa_a$ of each corotating black hole is analogous to the redshift $z_a$ of each corotating point particle, and more precisely
\beq\label{analogy}
	4 \mu_a \kappa_a \longleftrightarrow c_a z_a \, .
\eeq
Recall that the coefficient $c_a$ introduced in Eq.~\eqref{ploup} is nothing but the normalized horizon surface gravity of an isolated black hole of irreducible mass $\mu_a$ and horizon angular velocity $\omega_a$, say $4 \mu_a \bar{\kappa}_a$ [see Eqs.~\eqref{4mukappa_Kerr} and \eqref{ca}]. Therefore, the analogy \eqref{analogy} can equivalently be written as
\beq\label{analogy_bis}
	\frac{\kappa_a}{\bar{\kappa}_a} \longleftrightarrow z_a \, .
\eeq
Since the redshift of a particle satisfies $z_a \leqslant 1$, the analogy \eqref{analogy_bis} suggests that in corotating black hole binary systems the tidal interaction decreases the surface gravity of each black hole (with respect to its value in isolation). This agrees with the prediction of the perturbative calculation discussed in Sec.~\ref{subsec:BHPT}.

In the large separation limit, i.e. when $\Omega \to 0$, we know that $z_a \to 1$ for each corotating particle, while $\kappa_a \to \bar{\kappa}_a = 1/(4\mu_a)$ for each corotating black hole. Indeed, in that limit, the spin and proper rotation frequency of each body must vanish to maintain the corotation. To test the soundness of the analogy \eqref{analogy} beyond the large separation limit, we will compute the normalized black hole surface gravities $4 \kappa_a \mu_a$ in Sec.~\ref{sec:NR}, as well as the 4PN expansions of the particle's renormalized redshifts $c_a z_a$ in Sec.~\ref{sec:PN}, and compare those results, expressed a functions of the circular-orbit frequency $\Omega$, in Sec.~\ref{sec:compare}.

Interestingly, for black hole binaries with large mass ratios, the analogy \eqref{analogy_bis} is supported by a comparison, at the Hamiltonian level, of the expression for the surface gravity of a black hole tidally perturbed by a corotating moon, and that for the redshift of the most massive particle (say body $2$). Indeed, recalling Eqs.~\eqref{Dkappa} and \eqref{tic-tac}, we find the analogy
\beq\label{analogy_ter}
	\frac{\kappa}{\bar{\kappa}} = \frac{\partial \tilde{H}}{\partial M} \quad \longleftrightarrow \quad z_2 = \frac{\partial H_\text{F}}{\partial m_2} \, ,
\eeq
where we introduced the Hamiltonian $\tilde{H} \equiv M + z H$, including the mass $M$ of the background hole. The occurence of the moon's redshift $z$ in $\tilde{H}$ comes from the fact that the Hamiltonian $H$ of a test mass in Kerr is parameterized by the proper time $\tau$, while the Fokker-type two- body Hamiltonian $H_\text{F}$ is parameterized by the coordinate time $t$.

Similarly, by comparing the expression \eqref{Domega} for the perturbed angular velocity of the black hole to the corotation condition \eqref{cond_cor} with the formulas \eqref{tic-tac} for the point particle's proper rotation frequency, we note a formal analogy between $\bar{\omega}$ and $\omega_2$, namely
\beq\label{analogy_quad}
	\bar{\omega} = \biggl( \frac{\partial \tilde{H}}{\partial M} \biggr)^{-1} \biggl( \Omega - \frac{\partial \tilde{H}}{\partial S} \biggr) \quad \longleftrightarrow \quad \omega_2 = \biggl( \frac{\partial H_\text{F}}{\partial m_2} \biggr)^{-1} \biggl( \Omega - \frac{\partial H_\text{F}}{\partial S_2} \biggr) \, .
\eeq
In Secs.~\ref{sec:PN} and \ref{sec:BHPT} below, we will find that the large mass-ratio limits of the PN expansions of $z_2(\Omega)$ and $\omega_2(\Omega)$ are in full agreement with the PN expansions of the perturbative expressions for $(\kappa / \bar{\kappa})(\Omega)$ and $\bar{\omega}(\Omega)$, thus supporting the analogies \eqref{analogy_ter} and \eqref{analogy_quad}.

On the other hand, still for black hole binaries with large mass ratios, the analogy \eqref{analogy_bis} can be made rigorous for the smaller compact object (say body $1$). Indeed, using a method of matched asymptotic expansions, the author of Ref.~\cite{Po3.15} recently proved that  the horizon surface gravity $\kappa_1$ of a small black hole of mass $m$ in the tidal environment of a large black hole of mass $M \gg m$ is closely related to the redshift $z \equiv z_1$ of a point mass $m$ moving like a test particle in a certain \emph{effective} metric satisfying the linearized vacuum Einstein equation. More precisely, to linear order in the small mass ratio $q = m/M$, it can be shown that \cite{Po3.15}

\beq\label{Adam}
	\frac{\kappa_1}{\bar{\kappa}_1} = z \, .
\eeq
At this order of approximation, the smaller black hole can be taken to be nonrotating, such that $m = \mu_1$, $\bar{\kappa}_1 = 1/(4\mu_1)$ and $c_1 = 1$. Hence the result \eqref{Adam} proves that, at least to linear order in the mass ratio and for the smaller body $1$, Eq.~\eqref{analogy} is an identity, and not a mere analogy. See also the discussion in Ref.~\cite{Zi.al.16}.

\section{Numerical relativity}
\label{sec:NR}

In this section, we present our calculations of sequences of quasi-equilibrium initial data for corotating black hole binaries with mass ratios in the range $10:1-1:1$. In particular, we explain how the horizon surface gravity of each black hole is computed. The formulation used to construct the initial data is outlined in Sec.~\ref{subsec:formulation}, the dependence of our numerical data on the resolution is discussed in Sec.~\ref{subsec:convergence}, and our results are presented in Sec.~\ref{subsec:results}.

\subsection{Formulation}
\label{subsec:formulation}

The numerical results presented here are similar to those previosly published in \cite{Gr.al.02,Gr.10,Ur.al.12}. The basic properties of such data are as follows. We employ the 3+1 formalism of GR so that the four-dimensional line element reads
\beq
\ud s^2 = - \bigl( N^2 - B_i B^i \bigr) \, \ud t^2 + 2 B_i \, \ud x^i \ud t + \gamma_{ij} \, \ud x^i \ud x^j \, ,
\eeq
where $N$ is the lapse, $B^i$ the shift vector and $\gamma_{ij}$ the three-dimensional spatial metric. Spatial indices are lowered and raised by means of $\gamma_{ij}$ and its inverse $\gamma^{ij}$, so that e.g. $B_i \equiv \gamma_{ij} B^j$. Circularity of the orbit is enforced by imposing the existence of a helical Killing vector $k^\alpha$; recall Eq.~\eqref{helical_k}. From a practical point of view, the equations are written in the corotating frame, such that $k^\alpha = \left(\partial_t\right)^\alpha$ and the time derivatives of the various fields vanish.

Our corotation setup allows us to consider binary black hole systems with an \textit{exact} helical Killing symmetry, such that the apparent horizon is a Killing horizon, and the surface gravity is well defined over each component of the horizon. This should be contrasted to the recent work of Ref.\cite{Zi.al.16}, where the authors consider more generic and astrophysically realistic binary configurations, but do not have a well-defined (geometrical and unambiguous) notion of black hole surface gravity.

Far from the system, it is demanded that the spacetime be asymptotically flat. However, as discussed in Sec.~\ref{subsec:2BH}, this is not strictly compatible with the existence of {a global} helical symmetry. Indeed, if a true helical Killing vector existed, the system would have an infinite lifetime. This fact, coupled with the emission of gravitational waves would lead to diverging metric quantities, being the signature of this accumulated emission. There are two possible approaches to cure that problem. Either one can (to some extend) relax the Killing condition, or one needs to find a way to suppress the gravitational radiation. In this work, we opt for the latter solution. There are various ways to proceed. The most widely used is to impose that the solution be spatially conformally flat. The spatial metric $\gamma_{ij}$ is then simply related to the three-dimensional flat metric $f_{ij}$ through a conformal factor $\Psi$, namely
\beq
	\gamma_{ij} = \Psi^4 f_{ij} \, .
\eeq
Of course, the price to pay is that not all of Einstein's equations can be exactly satisfied.

With conformal flatness, the spacetime depends on five fields: the lapse $N$, the conformal factor $\Psi$ and the shift vector $B^i$. Those fields obey a subset of Einstein's equations, namely the trace of the evolution equation (\ref{traceevol}), the Hamiltonian constraint (\ref{hamilton}) and the momentum constraints (\ref{momentum}). The slicing (i.e. the choice of time) is done via maximal slicing, for which the trace of the extrinsic curvature tensor vanishes. The spacelike coordinates are fixed by the choice of corotating coordinates and the conformal flatness approximation. The resulting system of equations is then
\begin{subequations}\label{field}
	\begin{align}
		D_i D^i N &= - 2 D^i N \frac{D_i \Psi}{\Psi} + N \Psi^4 A_{ij} A^{ij} \, ,\label{traceevol} \\
		D_i D^i \Psi &= -\frac{\Psi^5}{8} A_{ij} A^{ij} \, , \label{hamilton} \\
		D_j D^j B^i + \frac{1}{3}D^i D_j B^j &= 2 A^{ij} \biggl( D_j N - 6 N \frac{D_j \Psi}{\Psi} \biggr) \, , \label{momentum}
	\end{align}
\end{subequations}
where $D$ denotes the covariant derivative compatible with the three-dimensional Euclidean metric $f_{ij}$, while $A^{ij} \equiv D^{\langle i} B^{j \rangle} / N$ is the conformal extrinsic curvature tensor.

The presence of two corotating black holes is enforced by imposing the following boundary conditions on two spheres corresponding to the apparent horizons:
\begin{subequations}
	\begin{align}
		N &= n_0 \, , \label{choicelapse} \\
		\partial_r \Psi + \frac{\Psi}{2r} &= - \frac{\Psi^3}{4} A_{ij} s^i s^j \, , \label{apparent} \\
		B^i &= \frac{n_0}{\Psi^2} \, s^i \, , \label{shifthor}
	\end{align}
\end{subequations}
where $s^i$ is the unit normal to each hole (with respect to the metric $f_{ij}$). Equation \eqref{choicelapse} is a free choice of the lapse on the boundaries of the physical domain; typically one uses $n_0 = 0.1$. This basically comes from the fact that maximal slicing is a differential gauge choice, defined uniquely up to the boundaries. This was discussed is some length in Refs.~\cite{Ca.al2.06, GoJa.06}. Equation \eqref{apparent} enforces that the two spheres are apparent horizons, and \eqref{shifthor} that the black holes are corotating. Asymptotic flatness is explicitly recovered by requesting that $N=1$, $\Psi=1$ and $B^i = \Omega \left(\partial_\varphi\right)^i$ at infinity. (Recall that we work in the corotating frame).

For each separation, the constant angular velocity $\Omega$ is determined by the numerical code to ensure the equality of the ADM and Komar masses:
\beq
	M_\text{ADM} = M_\text{K} \, .
\eeq
This condition is closely related to relativistic generalizations of the virial theorem \cite{GoBo.94,Fr.al.02}, and it has been used extensively to compute binary black hole configurations, e.g. in \cite{Gr.al.02, Gr.10, Ur.al.12}. The five elliptic equations \eqref{field} are solved by spectral methods implemented via the \texttt{Kadath} library. Bispherical coordinates are used and the algorithm is the same as the one already described in Refs.~\cite{Gr.10,UrTs.11}. The version of the code used for this work is more general in the sense that it can deal with black holes of different masses. When doing so, one needs to find the location of the rotation axis of the binary. This is done numerically and it is obtained from the fact that the total (ADM) linear momentum of the binary must vanish. The linear momentum can be computed as a surface integral at spatial infinity, and the condition can be rewritten as
\beq\label{cond_axis}
	\lim_{r \to \infty} \int A^{ij} \left(e_y\right)_i {\rm d}S_j = 0 \, ,
\eeq
where $\vec{e}_y$ is the unit vector (with respect to the flat metric) along the $y$-axis and $\ud S_i$ is the flat surface element at infinity. The various axes are defined as follows : the $x$-direction joins the center of the holes and the $z$-direction is the one of the rotation axis, so that the orbital plane is $z = 0$. The $y$-direction is then in the orbital plane, perpendicular to the line joining the holes. Due to the symmetries of the problem, only the $y$ component of the momentum is not trivially zero. It vanishes only when the axis is at the right location on the $x$-axis.

The coordinate radii of the holes are also unknowns. The sequences of corotating binaries are obtained by requesting that the irreducible mass of each hole is held fixed (see Sec.~\ref{subsubsec:ICO} below). The size of the numerical domain is allowed to vary and the numerical code finds the appropriate radii so that the masses have the desired values. More precisely, the irreducible masses are defined by $\mu_a \equiv \sqrt{A_a/(16\pi)}$, where $A_a$ is the surface area of the $a$\textsuperscript{th} hole, given by the following surface integral over a cross section $S_a$ of the corresponding horizon:
\beq\label{mirr}
A_a  = \int_{S_a} \! \Psi^4 \, {\rm d}S \, .
\eeq
Since, by construction, the horizons have a vanishing expansion, their areas $A_a$ do not depend on a particular choice of slicing \cite{As.al.00}. We computed quasi-equilibrium initial data for irreducible mass ratios $q_\mu \equiv \mu_1 / \mu_2 \in \{1,1/2,1/10\}$.

In order to compute the horizon surface gravity \eqref{0th_law} of each hole, we employ a formula given in terms of the 3+1 quantities. A suitable expression can be found, e.g., in Eq.~(10.10) of Ref.~\cite{GoJa.06}. There, the normal vector is normalized with respect to the spatial metric $\gamma_{ij}$, whereas the vector $s^i$ appearing in our Eqs.~\eqref{apparent} and \eqref{shifthor} is normalized with respect to the flat metric $f_{ij}$. It follows that the two normal vectors differ by a factor of $\Psi^2$. Moreover, the term proportional to $\ell^\alpha \nabla_\alpha \ln N$ in Eq.~(10.10) of Ref.~\cite{GoJa.06} is related to the evolution of the lapse function, and it vanishes for our quasi-stationary configurations. Taking all of that into account, the 3+1 expression for the surface gravity reduces to
\beq\label{kappa3p1}
\kappa_a = \frac{s^i D_i N}{\Psi^2} - N A_{ij} s^i s^j \Big|_{S_a}.
\eeq

\subsection{Effects of resolution}
\label{subsec:convergence}

Next, we explore the variations of some quantities of physical interest with respect to the numerical resolution. The three-dimensional physical space is split into several (typically a dozen) computational domains. In each domain the number of spectral coefficients is the same with respect to the three space dimensions. We denote by $N$ this number and compute binary configuration for $N=9$, $11$, $13$ and $15$ coefficients.

\begin{figure}[t!]
	\includegraphics[width=0.72\linewidth]{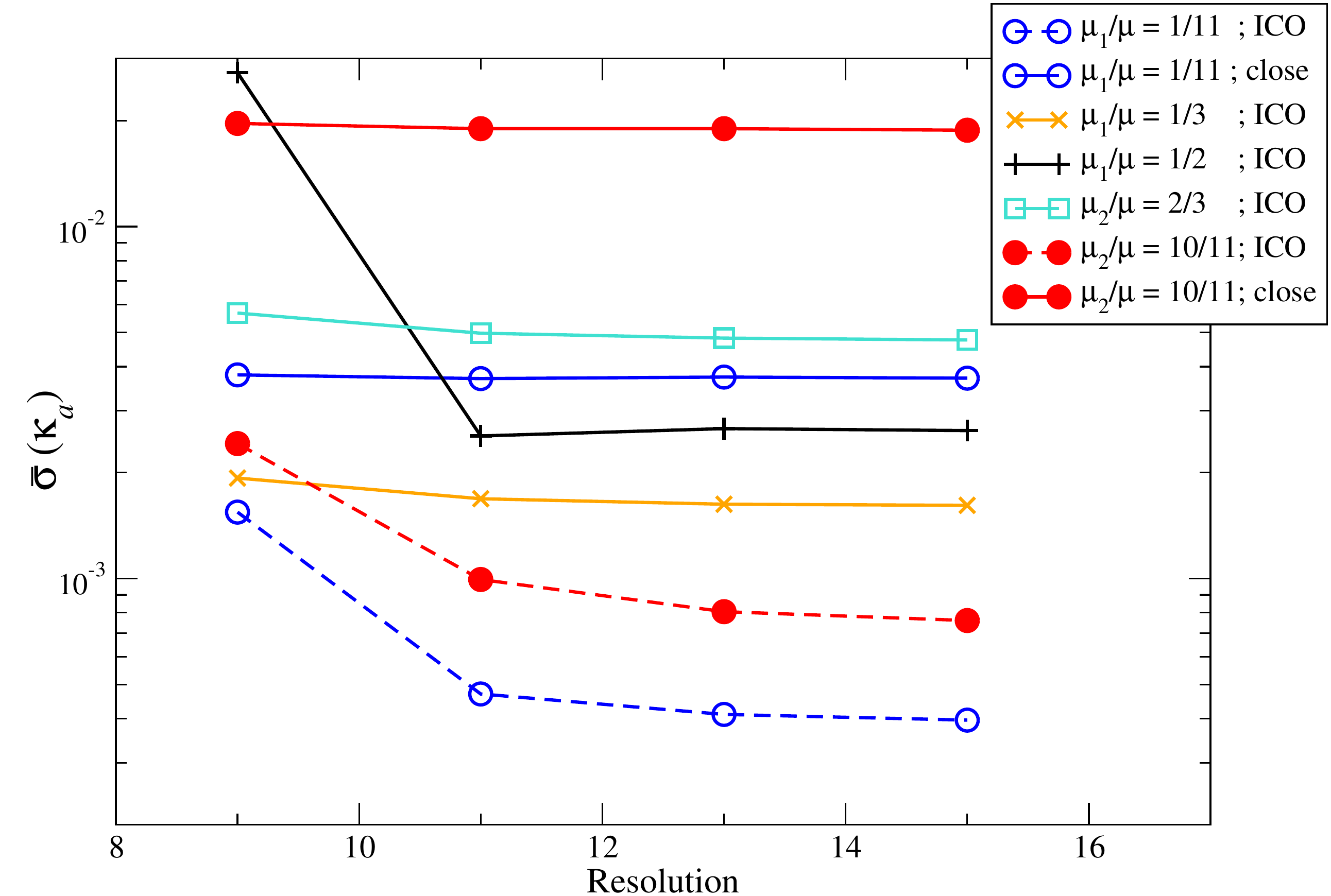}
	\caption{Normalized root-mean-square variation \eqref{var_kappa} of the surface gravity of each black hole, as a function of the number of points $N$ in each spatial dimension. Results for irreducible mass ratios $q_\mu \equiv \mu_1 / \mu_2 \in \{1,1/2,1/10\}$ are shown. For each unequal-mass configuration, two curves are plotted, one for each black hole, labelled by the fractional masses $\mu_1 / \mu$ and $\mu_2 / \mu$, with $\mu \equiv \mu_1 + \mu_2$ the total irreducible mass. Two different separations are displayed for $q_\mu=1/10$.}
	\label{fig:conv_var_kappa}
	\vspace{-0.2cm}
\end{figure}

The first test has to do with the zeroth law of binary black hole mechanics, being the fact that the horizon surface gravity \eqref{0th_law} of each hole is constant. To measure this we define the normalized root-mean-square variation (or relative standard deviation) of the surface gravity $\kappa_a$ of body $a$ by
\beq\label{var_kappa}
	\bar{\sigma}(\kappa_a) \equiv \sqrt{\av{\kappa_a^2}-\av{\kappa_a}^2} / \av{\kappa_a} \, .
\eeq
At each point one can compute $\kappa_a$ by Eq.~\eqref{kappa3p1}, and the various averages appearing in \eqref{var_kappa} are taken using all the collocation points located on the horizon of each black hole. Figure~\ref{fig:conv_var_kappa} shows $\bar{\sigma}(\kappa_a)$ as a function of $N$ for one configuration with equal masses, one with a mass ratio $q_\mu=1/2$ and two with $q_\mu=1/10$. All these configurations correspond to separations close to the innermost circular orbit (ICO), namely the configuration of minimum binding energy; see Sec.~\ref{subsubsec:ICO} below. For a mass ratio $q_\mu=1/10$ we show an additional configuration with a smaller separation, corresponding to an orbital frequency $\mu\Omega = 0.13$, where $\mu \equiv \mu_1 + \mu_2$ is the total irreducible mass. (In that case the ICO is located at $\mu\Omega = 0.025$; see Tab.~\ref{tab:ICO}). All the curves exhibit the same behavior, being the convergence of $\bar{\sigma}(\kappa_a)$ to a finite value as the resolution increases. It means that the error on the zeroth law is not coming from a numerical error, since that would depend on the resolution. Rather, it arises from the physical assumptions used when computing the binary configurations, such as our use of the CFC approximation, which implies that only a subset of Einstein's equations are solved.

\begin{figure}[t!]
	\includegraphics[width=0.72\linewidth]{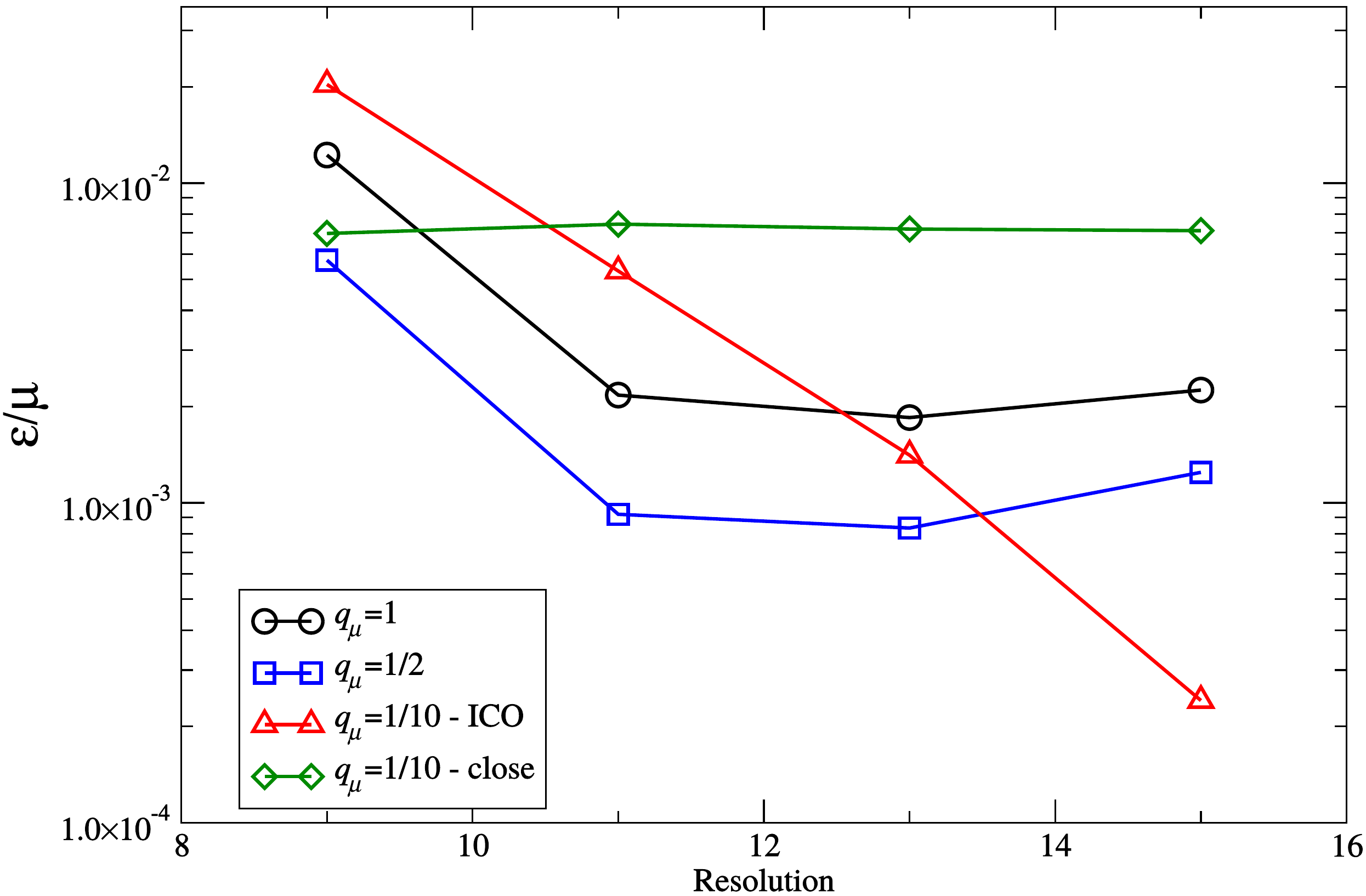}
	\caption{Error \eqref{varepsilon} on the first integral \eqref{1st_int_2BH} as a function of the numerical resolution, still for irreducible mass ratios $q_\mu \in \{1,1/2,1/10\}$. Two different separations are displayed for $q_\mu=1/10$.}
	\label{fig:conv_first}
	\vspace{-0.16cm}
\end{figure}

Another test is provided by the first integral relation \eqref{1st_int_2BH} associated with the variational first law \eqref{1st_law_2BH}. For each configuration, we compute the global quantities $M_\text{ADM}$, $J$ and $\Omega$, as well as the quasi-local quantities $\mu_a$ and $\av{\kappa_a}$. Figure \ref{fig:conv_first} shows the violation of this algebraic relation, as measured by the difference
\beq\label{varepsilon}
	\varepsilon \equiv M_\text{ADM} - 2\Omega J - \sum_a 4\mu_a^2 \av{\kappa_a} \, .
	\vspace{-0.3cm}
\eeq
More precisely, Fig. \ref{fig:conv_first} displays the variation of the dimensionless quantity $\varepsilon / \mu$ as a function of resolution, for the same configurations as the ones of Fig.~\ref{fig:conv_var_kappa}. The data for $q_\mu=1$ and $1/2$ exhibit a similar behavior, where the error saturates. The convergence with resolution is slightly less clean than for $\bar{\sigma}(\kappa_a)$, as the curves show a slight increase of the error for $N=15$. This increase is probably due to the difficulty in computing very accurately the global quantities $M_\text{ADM}$ and $J$. At this stage it is unclear how the error \eqref{varepsilon} would behave at even higher resolution, which is unfortunately out of reach of our current computational resources. Nevertheless, from Fig. \ref{fig:conv_first}, it is clear that the deviation from the first integral does not converge to zero as the resolution increases, but instead to a small but finite value. From this point of view it is very similar to what is observed for the zeroth law.

The data for $q_\mu=1/10$ at a separation close to the ICO (red triangles) exhibits a different behavior where no saturation is visible. The plausible explanation is that, in this case, the deviation from the first integral is so small that the error is still dominated by the effects of a finite resolution. This is consistent with results from Fig. \ref{fig:conv_var_kappa}, where the values of $\bar{\sigma}(\kappa_a)$ for $q_\mu=1/10$ at the ICO are significantly smaller that in all the other cases depicted. This explanation is also supported by data for a smaller separation (green diamonds). The black holes being closer, the errors induced by the CFC approximation are larger, as confirmed by Fig.~\ref{fig:conv_var_kappa}. In that case, the error on the first law is no longer dominated by resolution effects, such that the curve shows a plateau.

\subsection{Results}
\label{subsec:results}

Unless stated otherwise, all the results shown below correspond to the highest resolution, namely $N=15$ for irreducible mass ratios $q_\mu=1$ and $1/2$. For $q_\mu=1/10$, most of the results were computed while using $N=13$. A few configurations with $N=15$ were also computed to confirm that they were almost indistinguishable from the $N=13$ ones.

\subsubsection{Surface gravity and first integral}
\label{subsubsec:firstinteg}

\begin{figure}[b!]
	\includegraphics[width=0.75\linewidth]{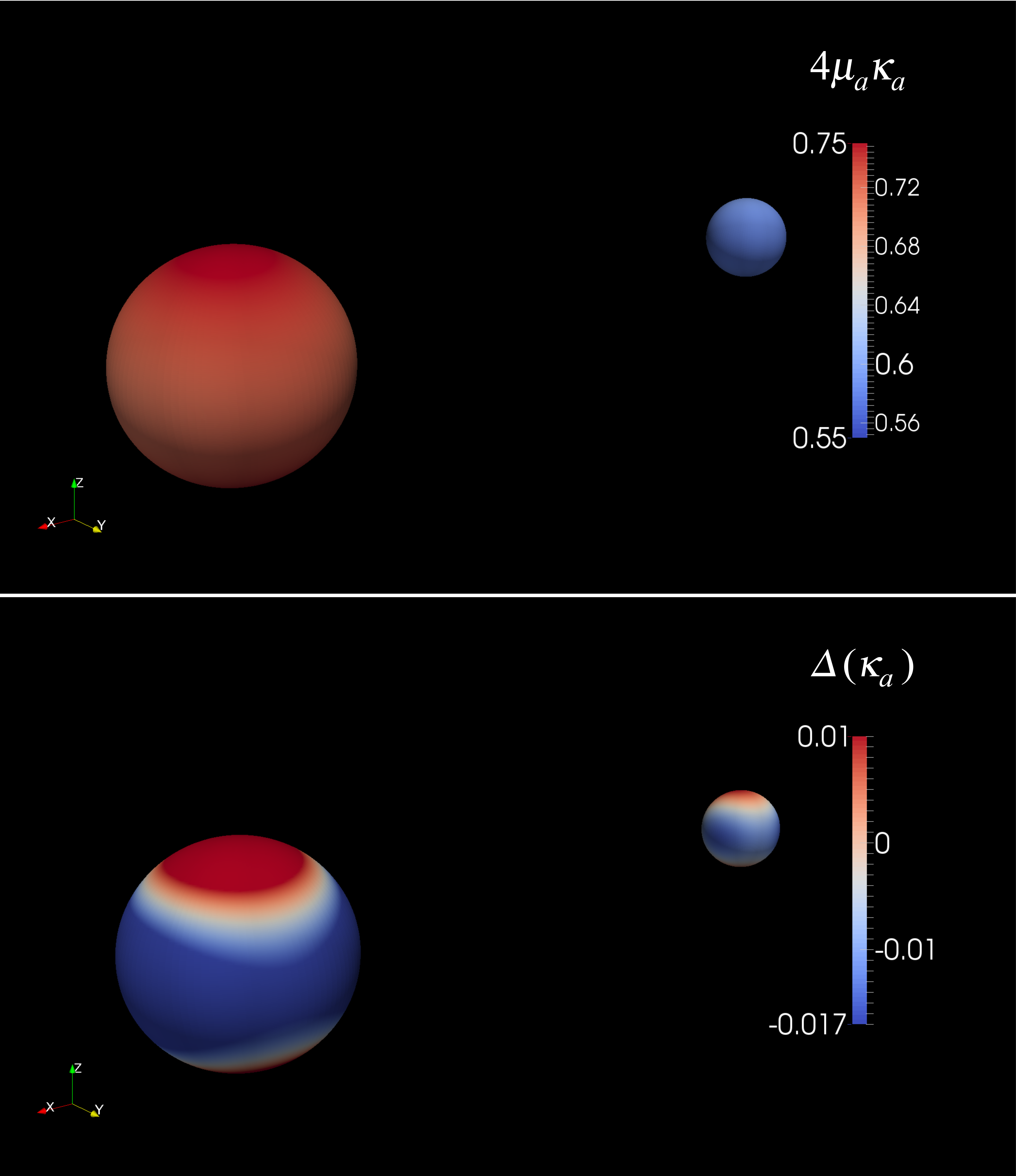}
	\caption{\textit{Top panel:} Normalized surface gravity $4\mu_a \kappa_a$ of each black hole for a corotating binary with mass ratio $q_\mu = \mu_1 / \mu_2 = 1/2$ and a circular-orbit frequency $\mu\Omega = 0.195$. \textit{Bottom panel:} Relative variations $\Delta(\kappa_a) = (\kappa_a - \av{\kappa_a}) / \av{\kappa_a}$ in horizon surface gravity for the same binary configuration.}
	\label{fig:visu_kappa}
\end{figure}

The top panel of Fig.~\ref{fig:visu_kappa} shows the normalized horizon surface gravity $4\mu_a \kappa_a$ of two corotating black holes with mass ratio $q_\mu = 1/2$, for a circular-orbit frequency $\mu\Omega = 0.195$. The horizons are depicted as spheres of constant coordinate radius. The bottom panel of Fig.~\ref{fig:visu_kappa} displays the relative variations in surface gravity,
\beq
	\Delta(\kappa_a) \equiv \frac{\kappa_a - \av{\kappa_a}}{\av{\kappa_a}} \, ,
\eeq
for the same configuration. Those reach at most a few percents, which shows that the zeroth law is satisfied to a high degree of accuracy, even for such a strong-field orbit. Indeed, for this configuration the ICO corresponds to an orbital frequency $\mu\Omega_\text{ICO} = 0.083$.

\begin{figure}[h!]
	\includegraphics[width=0.82\linewidth]{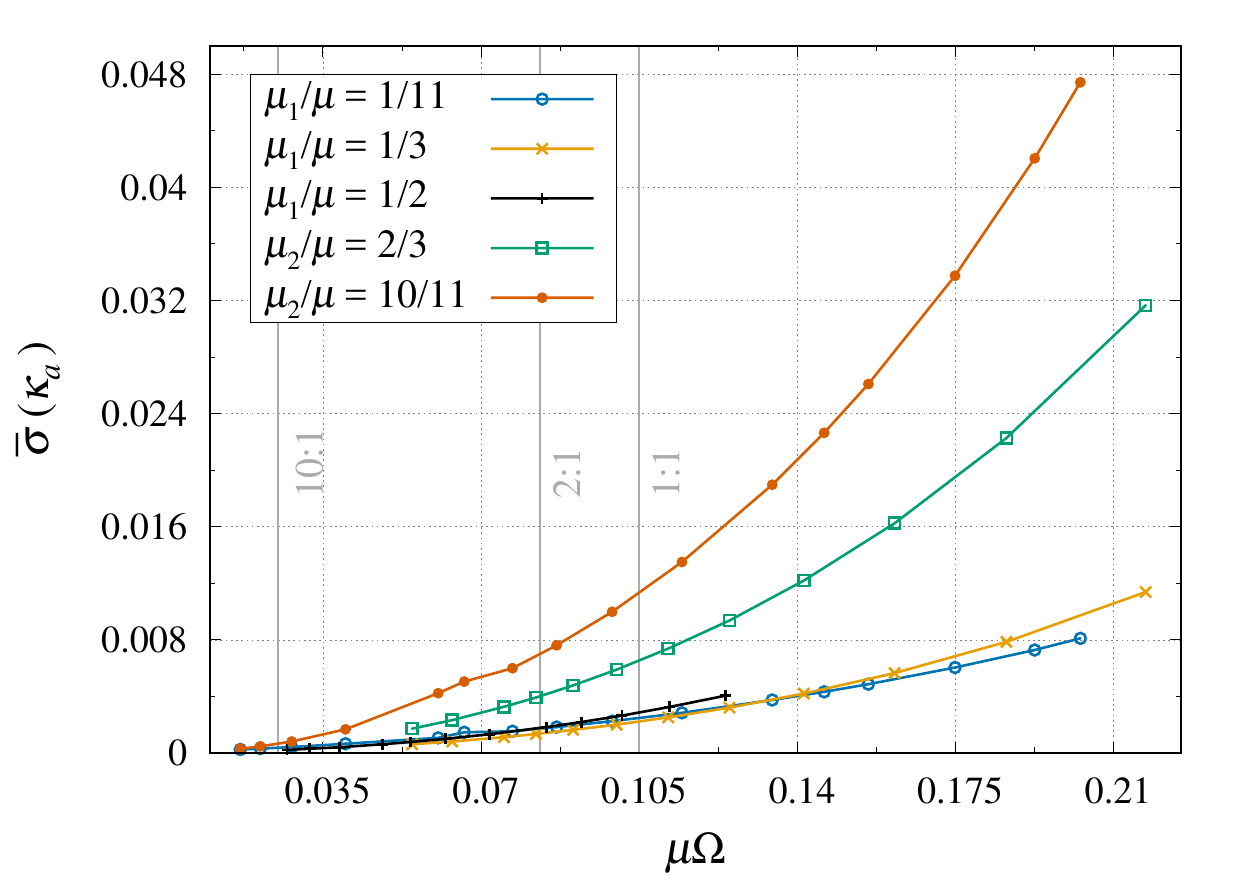}
	\caption{Normalized root-mean-square variation \eqref{var_kappa} of the horizon surface gravity of each black hole, as a function of the circular-orbit frequency $\mu\Omega$, for mass ratios $q_\mu \in \{1,1/2,1/10\}$. The gray vertical lines show the location of the innermost circular orbit for each mass ratio.}
	\label{fig:var_kappa}
\end{figure}

Figure \ref{fig:var_kappa} illustrates the deviation from the zeroth law. More precisely, it shows the value of $\bar{\sigma}(\kappa_a)$ as a function of the orbital frequency $\mu\Omega$, for mass ratios $q_\mu \in \{1,1/2,1/10\}$. The vertical gray lines show the location of the ICO in each case. For corotating orbits outside the ICO, the overall variation in surface gravity is always small, i.e., less than $0.3\%$, $0.4\%$ and $0.5\%$ for mass ratios $q_\mu = 1$, $1/2$ and $1/10$, respectively. However, for highly relativistic orbits inside the ICO, those variations can reach several percents, especially for the heaviest body.

Moreover, the variations displayed in Fig.~\ref{fig:var_kappa} are well above the numerical error level, and they show a monotonic increase as a function of frequency, for all configurations. This indicates that such variations are physical, in the sense that the zeroth law \eqref{0th_law} is not exactly satisfied. As discussed in Sec.~\ref{subsec:convergence}, this is to be expected because of the CFC approximation implemented in our initial data, which prevents us from having exact, helically symmetric solutions to the full set of Einstein field equations.

\begin{figure}[h!]
	\includegraphics[width=0.82\linewidth]{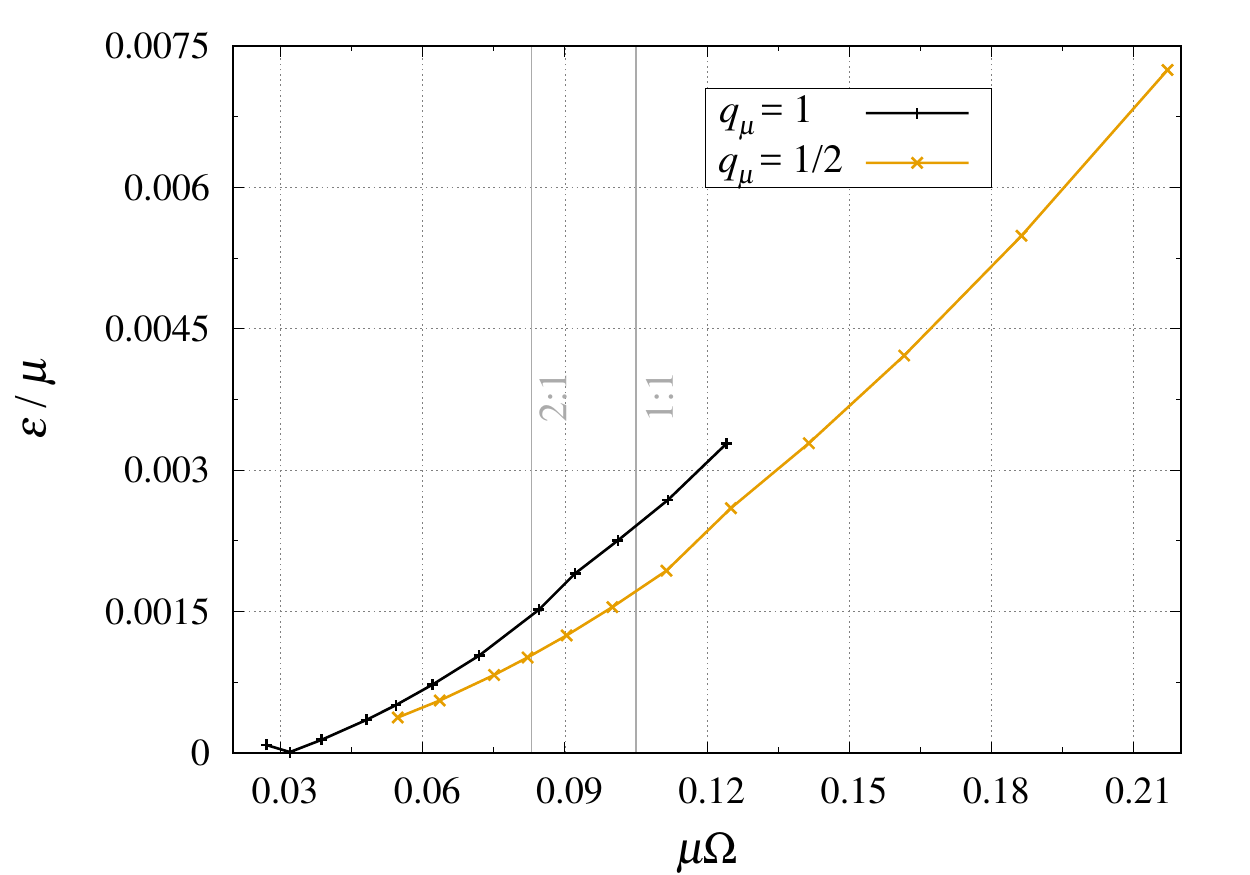}
	\caption{Violation \eqref{varepsilon} of the first integral relation \eqref{1st_int_2BH}, as a function of the orbital frequency $\mu\Omega$, for mass ratios $q_\mu = 1$ and $q_\mu = 1/2$.} 
	\label{fig:brk_1st_int}
\end{figure}

Figure \ref{fig:brk_1st_int} depicts the error $\varepsilon / \mu$ on the first integral, defined by \eqref{varepsilon}, as a function of the orbital frequency $\mu\Omega$, for mass ratios $q_\mu = 1$ and $1/2$. The first integral relation is satisfied at the $0.3\%$ and $0.1\%$ level at the ICO for $q_\mu = 1$ and $1/2$, and the dimensionless differences $(\varepsilon / \mu)(\mu\Omega)$ show a trend similar to that observed in Fig.~\ref{fig:var_kappa} for the normalized variations \eqref{var_kappa} in surface gravity. The data for $q_\mu = 1/10$ are too noisy to be displayed in Fig.~\ref{fig:brk_1st_int}, because in that case it gets challenging to extract the global quantities $M_\text{ADM}$ and $J$. The quantities \eqref{var_kappa} and \eqref{varepsilon} provide meaningful measures of the error (with respect to an exact solution of the full set of field equations) introduced by the CFC approximation.

\subsubsection{Innermost circular orbit}
\label{subsubsec:ICO}

The validity of the condition $\delta A_a = 0$, which we impose to compute our sequences of quasi-equilibrium initial data, is supported by the results from time evolutions of inspiralling black hole binaries. For instance, in the simplest case of nonspinning binaries, the total irreducible mass (defined as the sum of the irreducible masses of the apparent horizons) deviates from its initial value by only a few parts in $10^6$ over the entire inspiral phase \cite{Bo.al2.07,Bu.al.12}. This observed change is smaller than the numerical errors in the simulations, and analytical estimates \cite{Al.01,Po2.04,TaPo.08} suggest that the increase in irreducible mass due to tidal heating should be even smaller in this case. For equal-mass binaries with equal spins $\chi = +0.97$, however, the variation in total irreducible mass is noticeable over the inspiral, but the relative changes only increase from $10^{-6}$ to $1\%$ as the orbital velocity $v \equiv (M\Omega)^{1/3}$ ramps up from $0.3$ to $0.7$ \cite{Lo.al.12}. (Similar results hold for even larger spin values \cite{Sc.al.15}.) Since our corotating binaries have much smaller (dimensionless) spins, no larger than $0.2$ over our orbital frequency range, the approximation of constant irreducible masses is excellent.

Combining the condition $\delta A_a = 0$ with the first law of binary black hole mechanics \eqref{1st_law_2BH} implies that, along a given sequence,
\beq\label{dEdOmega}
	\frac{\ud E}{\ud \Omega} = \Omega \, \frac{\ud J}{\ud \Omega} \, ,
\eeq
where $E \equiv M_\text{ADM} - (\mu_1 + \mu_2)$ is the binding energy. While in principle we could determine the functions $E(\Omega)$ and $J(\Omega)$ and compute their slopes, in practice Eq.~\eqref{dEdOmega} only provides a weak test of the first law because of the sparseness of our data, and because of the limited accuracy of the measurements of the global quantities $M_\text{ADM}$ and $J$ for a given $\Omega$.

Along a given sequence of circular orbits, the orbit for which the binding energy $E(\Omega)$ is minimized is referred to as the minimum energy circular orbit (MECO), or more often as the innermost circular orbit (ICO). So according to \eqref{dEdOmega} the ICO frequency $\Omega_\text{ICO}$ satisfies
\beq
	\frac{\ud E}{\ud \Omega}\bigg|_{\text{ICO}} = \frac{\ud J}{\ud \Omega}\bigg|_{\text{ICO}} = 0 \, .
\eeq
The notion of ICO is closely related to that of innermost stable circular orbit (ISCO), which is defined by the vanishing of radial frequency associated with small eccentricity perturbations away from a given circular orbit. Indeed, for a test particle on a bound orbit around a Kerr black hole, the notions of ICO and ISCO are equivalent. More generally, when a meaningful conservative/dissipative split of the dynamics of a compact binary system can be defined,\footnote{This is known to be the case up to at least fourth order in the PN approximation \cite{Da.al.14,Be.al.16,Da.al.16,Be.al.17,Be.al2.17}, as well as for linear black hole perturbations \cite{HiFl.08}.} if the conservative dynamics derives from a Hamiltonian, then the notions of ICO and ISCO are equivalent \cite{Bu.al.03,ViFl.15,Fu.al.17}. When dissipative effects are included, however, the transition from inspiral to plunge does not occure at a particular, sharp frequency (the ICO frequency), but is replaced instead by a smooth transition over a frequency range \cite{OrTh.00,BuDa.00,BaSa.09}. The notion of ICO has thus been used extensively as a useful reference point for comparisons between different analytical approximation methods and numerical techniques, e.g. in \cite{Da.al3.00,Pf.al.00,Gr.al.02,Bl.02,Da.al.02,CoPf.04,Ca.al2.06,Fa.11,Fa2.11}.

Table \ref{tab:ICO} shows the numerical values of the circular-orbit frequency $\mu\Omega$, the reduced binding energy $E/\mu$ and the dimensionles angular momentum $J/\mu^2$ of the ICO for corotating binaries with mass ratios $q_\mu \in \{1,1/2,1/10\}$, as computed using the CFC and PN approximations. The PN predictions were computed from the 4PN-accurate expressions for the binding energy $E(\Omega)$ and angular momentum $J(\Omega)$ of binary systems of corotating points particles, as given by the formulas \eqref{E-J_nonspin}--\eqref{E-J_corot} in App.~\ref{app:E-J}. The accuracy on the CFC quantities is asserted from convergence studies for the same resolutions as in Figs.~\ref{fig:conv_var_kappa} and \ref{fig:conv_first}. Convergence gets slower as the mass ratio decreases. For $q_\mu = 1/10$ our highest available resolution is not high enough to extract the binding energy in a satisfactory manner. While for all mass ratios the agreement between the PN and NR predictions improves with increasing order, the PN series does not appear to converge precisely towards the CFC results, with systematic relative differences of a few percents, consistent with previous results (see for instance Ref.~\cite{Da.al.02}). This is most likely due to the fact that those two approximations are of a different nature.

\vspace{-0.1cm}
\begin{center}
	\begin{table}[h!]
		\begin{tabular}{llcl|lcl|lcl}
			\toprule
			\multicolumn{10}{c}{\hspace{1.5cm}$q_\mu=1$\hspace{3cm}$q_\mu=1/2$\hspace{3cm}$q_\mu=1/10$} \\
			\cmidrule(r){2-10}
				& $\mu\Omega_\text{ICO}$ & $E_\text{ICO}/\mu$ & $J_\text{ICO}/\mu^2$ & $\mu\Omega_\text{ICO}$ & $E_\text{ICO}/\mu$ & $J_\text{ICO}/\mu^2$ & $\mu\Omega_\text{ICO}$ & $E_\text{ICO}/\mu$ & $J_\text{ICO}/\mu^2$ \\
			\midrule
			1PN & $0.522$  & $-0.04054$ & $0.6208$ & $0.525$  & $-0.03614$ & $0.5510$ & $0.537$  & $-0.01365\;\,$ & $0.2034$ \\
			2PN & $0.0809$ & $-0.01446$ & $0.8819$ & $0.0684$ & $-0.01159$ & $0.8176$ & $0.0250$ & $-0.002282$ & $0.4010$ \\
			3PN & $0.0984$ & $-0.01557$ & $0.8589$ & $0.0781$ & $-0.01230$ & $0.8020$ & $0.0255$ & $-0.002301$ & $0.3998$ \\
			4PN & $0.0998$ & $-0.01572$ & $0.8597$ & $0.0787$ & $-0.01230$ & $0.8023$ & $0.0255$ & $-0.002303$ & $0.3997$ \\
			CFC & $0.105$  & $-0.0164$  & $0.843$  & $0.083$  & $-0.0126$  & $0.789$  & $0.025$  & $---$ & $0.4(5)$ \\
			\bottomrule
		\end{tabular}
		\caption{Numerical values of the innermost circular orbit (ICO) frequency $\mu\Omega_\text{ICO}$, reduced binding energy $E_\text{ICO}/\mu$ and dimensionless angular momentum $J_\text{ICO}/\mu^2$, for corotating binaries with mass ratios $q_\mu \in \{1,1/2,1/10\}$, as computed using the CFC and PN approximations. The numerical data for $q_ \mu=1/10$ are difficult to compute due to a limited resolution (see also Sec. \ref{subsubsec:firstinteg}).}
		\label{tab:ICO}
	\end{table}
\end{center}
%

\section{Post-Newtonian theory}
\label{sec:PN}

In this section we shall derive, in the context of the PN approximation, the 4PN-accurate expressions of the redshifts for binary systems of corotating spinning point masses. This requires the knowledge of the redshifts up to 4PN order for nonspinning systems (Sec.~\ref{subsec:PN_NS}), that of the spin-orbit contributions up to next-to-leading order (Sec.~\ref{subsec:PN_SO}), and the application of the corotation condition \eqref{cond_cor} to express the spins as functions of the circular-orbit frequency (Sec.~\ref{subsec:PN_cor}), from which the renormalized redshifts can be derived (Sec.~\ref{subsec:PN_cor2}).

\subsection{Redshifts for nonspinning binaries}\label{subsec:PN_NS}

For nonspinning (NS) binaries, the relations $z^\text{ns}_a(\Omega)$ were computed up to 2PN and later to 3PN order in Refs.~\cite{De.08,Bl.al.10}. The leading and next-to-leading order logarithmic contributions at 4PN and 5PN orders, which originate from gravitational-wave tails, were then obtained in Refs.~\cite{Bl.al2.10,Da.10,Le.al.12}. Those results were all derived starting from the definition of the redshift in terms of the spacetime metric $g_{\alpha\beta}(x)$ evaluated at the coordinate location $x^\alpha = y_a^\alpha$ of the particle $a$, namely
\beq
	z_a = \frac{\ud \tau_a}{\ud t} = \left[ - g_{\alpha\beta}(y_a) v_a^\alpha v_a^\beta \right]^{1/2} .
\eeq
Moreover, the nonlogarithmic 4PN terms were also computed, from the known 4PN contribution to the circular-orbit binding energy \cite{JaSc.13,BiDa.13} combined with the first law of mechanics \eqref{1st_law_2PP} for binary systems of nonspinning point masses \cite{Le.al.12,Le.15,BlLe.17}. All these results are valid for arbitrary mass ratios. Up to 4PN order, we have
\begin{align}\label{z1NS}
	z_1^\text{ns} &= 1 + \left( - \frac{3}{4} - \frac{3}{4} \Delta + \frac{\nu}{2} \right) x + \left( - \frac{9}{16} - \frac{9}{16} \Delta - \frac{\nu}{2} - \frac{1}{8} \Delta \, \nu + \frac{5}{24} \nu^2 \right) x^2 \nonumber \\ &\qquad\! + \left( - \frac{27}{32} - \frac{27}{32} \Delta - \frac{\nu}{2} + \frac{19}{16} \Delta \, \nu - \frac{39}{32} \nu^2 - \frac{1}{32} \Delta \, \nu^2 + \frac{\nu^3}{16} \right) x^3 \nonumber \\ &\qquad\! + \left( - \, \frac{405}{256} - \frac{405}{256} \Delta + \left[ \frac{38}{3} - \frac{41}{64} \pi^2 \right] \nu + \left[ \frac{6889}{384} - \frac{41}{64} \pi^2 \right] \Delta \, \nu \right. \nonumber \\ &\qquad\qquad\!\! \left. + \left[ - \frac{3863}{576} + \frac{41}{192} \pi^2 \right] \nu^2 - \frac{93}{128} \Delta \, \nu^2 + \frac{973}{864} \nu^3 - \frac{7}{1728} \Delta \, \nu^3 + \frac{91}{10368} \nu^4 \right) x^4 \nonumber \\ &\qquad\! + \left( - \, \frac{1701}{512} - \, \frac{1701}{512} \Delta + \left[ - \frac{329}{15} + \frac{1291}{1024} \pi^2 + \frac{64}{5} \gamma_\text{E} + \frac{32}{5} \ln{(16x)} \right] \nu  \right. \nonumber \\ &\qquad\qquad\!\! + \left[ - \frac{24689}{3840} + \frac{1291}{1024} \pi^2 + \frac{64}{5} \gamma_\text{E} + \frac{32}{5} \ln{(16x)} \right] \Delta \, \nu + \left[ - \frac{71207}{1536} + \frac{451}{256} \pi^2 \right] \Delta \, \nu^2 \nonumber \\ &\qquad\qquad\!\! + \left[ - \frac{1019179}{23040} + \frac{6703}{3072} \pi^2 + \frac{64}{15} \gamma_\text{E} + \frac{32}{15} \ln{(16x)} \right] \nu^2 + \left[ \frac{356551}{6912} - \frac{2255}{1152} \pi^2 \right] \nu^3 \nonumber \\ &\qquad\qquad\!\! \left. + \,\, \frac{43}{576} \, \Delta \, \nu^3 - \frac{5621}{41472} \, \nu^4 + \frac{55}{41472} \, \Delta \, \nu^4 - \frac{187}{62208} \, \nu^5 \right) x^5 + o(x^5) \, ,
\end{align}
where $x \equiv (m \Omega)^{2/3}$ is the small frequency-related PN parameter, with $m \equiv m_1 + m_2$ the total mass, while $\nu \equiv m_1 m_2 / m^2$ is the symmetric mass ratio and $\Delta \equiv (m_2 - m_1) / m = \sqrt{1 - 4 \nu}$ is the reduced mass difference. (We assume that $m_1 \leqslant m_2$.) The expression for $z_2^\text{ns}$ is easily deduced by setting $\Delta \to -\Delta$ in Eq.~\eqref{z1NS}. Note the occurence of a logarithmic running and of Euler's constant $\gamma_\text{E}$ at 4PN order.

\subsection{Spin-orbit contributions to the redshifts and spin precession frequencies}\label{subsec:PN_SO}

Since our goal is to compute the redshifts for corotating spinning point-particle binaries, we need to account for the spin contributions to the redshifts. At the relative 4PN accuracy, however, it will prove sufficient to include the leading order 1.5PN and the next-to-leading order 2.5PN spin-orbit (SO) contributions. Up to next-to-leading order, the SO contributions to the redshift of body 1 read \cite{Bl.al.13}
\begin{align}\label{z1SO}
	z_1^\text{so} &= \left[ \left( - \frac{1}{3} + \frac{\Delta}{3} + \frac{2}{3} \nu \right) \nu \, \chi_1 + \left( 1 + \Delta - \frac{17}{6} \nu - \frac{5}{6} \Delta \, \nu + \frac{2}{3} \nu^2 \right) \chi_2 \right] x^{5/2} \nonumber \\ &+ \biggl[ \left( \frac{1}{2} - \frac{\Delta}{2} + \frac{19}{18} \nu - \frac{19}{18} \Delta \, \nu - \frac{\nu^2}{9} \right) \nu \, \chi_1 + \biggl( \frac{3}{2} + \frac{3}{2} \Delta - \frac{17}{3} \nu - \frac{8}{3} \Delta \, \nu \nonumber \\ &\qquad + \frac{179}{36} \nu^2 + \frac{41}{36} \Delta \, \nu^2 - \frac{\nu^3}{9} \biggr) \, \chi_2 \biggr] \, x^{7/2} + \mathcal{O}(x^{9/2}) \, ,
\end{align}
where $\chi_a \equiv S_a / m_a^2$ are the dimensionless spins of the two compact objects. The expression for $z_2^\text{so}$ is easily deduced by setting $\Delta \to - \Delta$ and $\chi_1 \leftrightarrow \chi_2$ in Eq.~\eqref{z1SO}. We will also need the expressions for the spin precession frequencies $\Omega_a$ as functions of the circular-orbit frequency $\Omega$. These were computed up to next-to-leading order in \cite{Bl.al.13}, from the spin-orbit part of the binary canonical Hamiltonian, and up to next-to-next-to-leading order in Refs.~\cite{Bo.al.13} and \cite{Do.al.14}. For body $1$, we have
\begin{align}\label{Omega_a}
	\Omega_1 &= \Omega \, \biggl\{ \left( \frac{3}{4} + \frac{3}{4} \Delta + \frac{\nu}{2} \right) x + \left( \frac{9}{16} + \frac{9}{16} \Delta + \frac{5}{4} \nu - \frac{5}{8} \Delta \, \nu - \frac{\nu^2}{24} \right) x^2 \nonumber \\ &\qquad\quad  + \left( \frac{27}{32} + \frac{27}{32} \Delta + \frac{3}{16} \nu - \frac{39}{8} \Delta \, \nu - \frac{105}{32} \nu^2 + \frac{5}{32} \Delta \, \nu^2 - \frac{\nu^3}{48} \right) x^3 + \mathcal{O}(x^4) \biggr\} \, .
\end{align}

Now, when reducing to corotating binaries, we will consider the sum $z_a = z_a^\text{ns} + z_a^\text{so} + \calO(\chi^2)$ and impose the corotation condition \eqref{cond_cor} to express the proper rotation frequencies $\omega_a$, and then the dimensionless spins $\chi_a$, as functions of the circular-orbit frequency $\Omega$. Importantly, while doing so, it is not necessary to account for spin-spin (or any higher order in the spins) contributions to the redshifts $z_a$. Indeed, spin-spin effects occure at the leading 2PN order, but since the spins themselves contribute at the leading 1.5PN order for corotating binaries [see Eq.~\eqref{chi_a_corot} below], these spin-spin interactions are pushed to 5PN order, which is beyond the accuracy of our calculations.

Beware, however, that the leading spin-spin couplings in the Hamiltonian dynamics contribute terms at the relative 0.5PN order beyond the leading term in the expansion \eqref{Omega_a} of the spin precession frequencies, which then contribute at the relative 2PN order for corotating binaries. Since the first law \eqref{1st_law_2PPspin} has only been established to linear order in the spins, we do not include such quadratic-in-spin effects here. Including these would not impact our final result for the renormalized redshifts, Eq.~\eqref{c1z1} below.

\subsection{Proper rotation frequencies and spins for corotating binaries}\label{subsec:PN_cor}

\hspace{-0.02cm}We now consider binary systems of corotating spinning point particles. Such systems are fully characterized by the component masses $m_a$, or equivalently by the irreducible masses $\mu_a$, and the angular frequency $\Omega$ of the orbit. In particular, the redshifts $z_a$, the spins $S_a$, the spin precession frequencies $\Omega_a$ and the proper rotation frequencies $\omega_a$ should all be functions of these three parameters. As discussed in Sec.~\ref{sec:laws} above, the condition for corotation is given by Eq.~\eqref{cond_cor}, which is equivalent to
\beq\label{omega_a}
	\omega_a(\Omega) = z^{-1}_a(\Omega) \left[ \Omega - \Omega_a(\Omega) \right] ,
\eeq
where the redshifts $z_a$ and spin precession frequencies $\Omega_a$ are functions of the binary's orbital frequency $\Omega$ (and the component masses $m_a$) only. Substituting for Eqs.~\eqref{z1NS} and \eqref{Omega_a} into \eqref{omega_a} yields the following 3PN-accurate expression for the proper rotation frequency of the largest body:
\beq\label{omega_a_PN}
	\omega_2 = \Omega \left\{ 1 - \nu \, x - \nu \left( \frac{3}{2} - \frac{\nu}{3} \right) x^2 - \nu \left( \frac{11}{8} + 2 \Delta - 5 \nu \right) x^3 + \mathcal{O}(x^4) \right\} .
\eeq
Here, it is sufficient to use the expression \eqref{z1NS} for nonspinning bodies because the spin-orbit terms \eqref{z1SO} contribute at leading relative 4PN order in Eq.~\eqref{omega_a_PN}. Interestingly, even though $z_1 \neq z_2$ and $\Omega_1 \neq \Omega_2$, we notice that $\omega_1 = \omega_2$ up to relative 2PN order \cite{Bl.al.13}. In the Newtonian limit $x \to 0$ and in the extreme mass-ratio limit $\nu \to 0$, we recover $\omega_a = \Omega$, in agreement with physical intuition. In Sec.~\ref{subsec:spin}, we will demonstrate how using the 3PN-accurate result \eqref{omega_a_PN} for equal-mass binaries (for which $\nu = \tfrac{1}{4}$ and $\Delta = 0$), instead of the Newtonian result $\omega_a = \Omega$, improves significantly the measurements of quasi-local spins performed in Ref.~\cite{Ca.al2.06}.

We may then compute the dimensionless spins $\chi_a = S_a / m_a^2$ as functions of the circular-orbit frequency $\Omega$ by combining Eqs.~\eqref{ma} and \eqref{Sa}, in which we substitute the formula \eqref{omega_a_PN} for the proper rotation frequencies. Up to next-to-leading order, which is sufficient for our purposes, we obtain
\beq\label{chi_a_corot}
	\chi_a = 4 \mu_a \omega_a \sqrt{1 - 4 (\mu_a \omega_a)^2} =  2 y^{3/2} \left( 1 \mp \Gamma \right) \left\{ 1 - \eta \, y + \mathcal{O}(y^2) \right\} ,
\eeq
where $y \equiv (\mu \Omega)^{2/3}$ is a frequency-related PN parameter, with the frequency adimensionalized using the total irreducible mass $\mu \equiv \mu_1 + \mu_2$, while $\eta \equiv \mu_1 \mu_2 / \mu^2$ is the symmetric irreducible mass ratio and $\Gamma \equiv (\mu_2 - \mu_1) / \mu = \sqrt{1 - 4 \eta}$ the irreducible mass difference. The parameters $(\mu,\eta,\Gamma,y)$ are analogous to the parameters $(m,\nu,\Delta,x)$ used above, but are defined in terms of the irreducible masses $\mu_a$, rather than the ordinary masses $m_a$. For corotating systems, Eqs.~\eqref{ma} and \eqref{omega_a_PN} can be combined to give
\begin{subequations}
	\begin{align}
		m &= \mu \left\{ 1 + \left( 2 - 6\eta \right)  \left( 1 - 2 \eta \, y \right) y^3 + \eta \biggl( -6 + \frac{64}{3} \eta - 10 \eta^2 \biggr) y^5 + o(y^5) \right\} , \\
		x &= y \left\{ 1 + \biggl( \frac{4}{3} - 4\eta \biggr) \! \left( 1 - 2 \eta \, y \right) y^3 + o(y^4) \right\} , \\
		\nu &= \eta \, \Bigl\{ 1 - \left( 2 - 8\eta \right) \left( 1 - 2 \eta \, y \right) y^3 + o(y^4) \Bigr\} \, , \\
		\Delta &= \Gamma \, \Bigl\{ 1 +4 \eta \, y^3 - 8 \eta^2 y^4 + o(y^4) \Bigr\} \, .
	\end{align}
\end{subequations}
As anticipated in Sec.~\ref{subsec:PN_SO} above, for corotating binaries the spins contribute at the leading 1.5PN order. The expressions \eqref{chi_a_corot} for the dimensionless spins $\chi_a$ are to be substituted into the leading and next-to-leading SO contributions \eqref{z1SO} to the redshift, which then contribute at the 3PN and 4PN orders, respectively.

\subsection{Renormalized redshifts for corotating binaries}\label{subsec:PN_cor2}

Next, to compute the renormalized redshifts $c_a z_a$ for corotating spinning point particles, we also need to control the 4PN expansions of the response coefficients $c_a$ as functions of the circular-orbit frequency $\Omega$. By substituting Eq.~\eqref{omega_a_PN} into \eqref{ca}, we obtain the following 4PN result for body $1$:
\begin{align}\label{c1}
	c_1 &= 1 + \left( -3 + 3 \Gamma + 6 \eta \right) y^3 + \eta \left( 6 - 6 \Gamma - 12 \eta \right) y^4 \nonumber \\ &\qquad + \eta \left( 9 - 9 \Gamma - 23 \eta + 5 \Gamma \, \eta + 10 \eta^2 \right) y^5 + o(y^5) \, .\end{align}
Notice the absence of contributions at $\calO(y)$ and at $\calO(y^2)$. In the test-particle limit $\mu_1 \to 0$, $\Gamma \to 1$ and $\eta \to 0$, such that $c_1 = 4\mu_1\bar{\kappa}_1 = 1 + \calO(q^2)$ and $c_2 = 4\mu_2\bar{\kappa}_2 = 1 - 6y^3 + o(y^5) + \calO(q)$.

Finally, by adding the nonspinning and spin-orbit contributions to the redshift, Eqs.~\eqref{z1NS} and \eqref{z1SO}, and using the expression \eqref{chi_a_corot} for the corotating spins, we obtain after multiplying by Eq.~\eqref{c1} and expanding in powers of the small PN parameter:
\begin{align}\label{c1z1}
	c_1 z_1 &= 1 + \left( - \frac{3}{4} - \frac{3}{4} \Gamma + \frac{\eta}{2} \right) y + \left( - \frac{9}{16} - \frac{9}{16} \Gamma - \frac{\eta}{2} - \frac{1}{8} \Gamma \, \eta + \frac{5}{24} \eta^2 \right) y^2 \nonumber \\ &\qquad\! + \left( - \frac{123}{32} + \frac{69}{32} \Gamma + \frac{11}{2} \eta + \frac{19}{16} \Gamma \, \eta - \frac{39}{32} \eta^2 - \frac{1}{32} \Gamma \, \eta^2 + \frac{\eta^3}{16} \right) y^3 \nonumber \\ &\qquad\! + \left( \frac{363}{256} + \frac{363}{256} \Gamma + \left[ \frac{23}{3} - \frac{41}{64} \pi^2 \right] \eta + \left[ \frac{1129}{384} - \frac{41}{64} \pi^2 \right] \Gamma \, \eta \right. \nonumber \\ &\qquad\qquad\!\! \left. + \left[ - \frac{983}{576} + \frac{41}{192} \pi^2 \right] \eta^2 - \frac{93}{128} \Gamma \, \eta^2 + \frac{973}{864} \eta^3 - \frac{7}{1728} \Gamma \, \eta^3 + \frac{91}{10368} \eta^4 \right) y^4 \nonumber \\ &\qquad\! + \left( \frac{603}{512} + \frac{603}{512} \Gamma + \left[ - \frac{494}{15} + \frac{1291}{1024} \pi^2 + \frac{64}{5} \gamma_\text{E} + \frac{32}{5} \ln{(16y)} \right] \eta  \right. \nonumber \\ &\qquad\qquad\!\! + \left[ - \frac{147569}{3840} + \frac{1291}{1024} \pi^2 + \frac{64}{5} \gamma_\text{E} + \frac{32}{5} \ln{(16y)} \right] \Gamma \, \eta + \left[ - \frac{33767}{1536} + \frac{451}{256} \pi^2 \right] \Gamma \, \eta^2 \nonumber \\ &\qquad\qquad\!\! + \left[ - \frac{703339}{23040} + \frac{6703}{3072} \pi^2 + \frac{64}{15} \gamma_\text{E} + \frac{32}{15} \ln{(16y)} \right] \eta^2 + \left[ \frac{169351}{6912} - \frac{2255}{1152} \pi^2 \right] \eta^3 \nonumber \\ &\qquad\qquad\!\! \left. + \,\, \frac{43}{576} \, \Gamma \, \eta^3 - \frac{5621}{41472} \, \eta^4 + \frac{55}{41472} \, \Gamma \, \eta^4 - \frac{187}{62208} \, \eta^5 \right) y^5 + o(y^5) \, .
\end{align}
The result for body $2$, such that $\mu_2 \geqslant \mu_1$, is straightforwardly deduced by setting $\Gamma \to - \Gamma$.

Anticipating a comparison to the perturbative results to be derived in Sec.~\ref{sec:BHPT}, we consider the PN expansion \eqref{c1z1}, as well as that for body $2$, in the limit $q_\mu \equiv \mu_1 / \mu_2 \ll 1$ of large mass ratios. To do so, it is convenient to express the final results in terms of the frequency-related parameter $u \equiv (\mu_2 \Omega)^{2/3}$, rather than $y = (\mu \Omega)^{2/3} = u \, (1+q_\mu)^{2/3}$. We find
\begin{subequations}
	\begin{align}
		c_1 z_1 &= 1 - \frac{3}{2} u - \frac{9}{8} u^2- \frac{27}{16} u^3 + \frac{363}{128} u^4 + \frac{603}{256} u^5 + q_\mu \, \biggl( u - u^2 - u^3 + \biggl[ \frac{46}{3} - \frac{41}{32} \pi^2 \biggr] u^4 \nonumber \\ &\qquad + \biggl[ - \frac{988}{15} + \frac{1291}{512} \pi^2 + \frac{128}{5} \gamma_\text{E} +\frac{64}{5} \ln{(16u)} \biggr] u^5 \biggr) + o(u^5,q_\mu) \, , \label{c1z1exp} \\
		c_2 z_2 &= 1 - 6 u^3 - q_\mu \left( u + \frac{3}{2} u^2 + \frac{27}{8} u^3 - \frac{121}{16} u^4 - \frac{1005}{128} u^5 \right) + o(u^5,q_\mu) \, . \label{c2z2exp}
	\end{align}
\end{subequations}
%

\section{Black hole perturbation theory}
\label{sec:BHPT}

In this section, we use the results of Ref.~\cite{GrLe.13} to express the surface gravity of a spinning black hole perturbed by a corotating moon in terms of the circular-orbit frequency (Sec.~\ref{subsec:kappa}). We then use the result \eqref{Adam} in Sec.~\ref{subsec:z-kappa}, together with recent gravitational self-force results, to compute the surface gravity of the orbiting moon itself (Sec.~\ref{subsec:kappa_moon}). Finally, we rely on the symmetry properties of the PN expression \eqref{c1z1} for the renormalized redshifts to introduce a physically-motivated ``mass-ratio rescaling'' of the perturbative expressions (Sec.~\ref{subsec:sym}).

\subsection{Surface gravity of a black hole with a corotating moon}\label{subsec:kappa}

The formulas \eqref{kappa_Kerr} and \eqref{Dkappa2} are expressed in terms of the spin $\chi$ and angular velocity $\bar{\omega}$ of the background Kerr black hole. Instead, we wish to express the surface gravity $\kappa_2 \equiv \kappa = \bar{\kappa} + \calD \kappa + \mathcal{O}(q^2)$ of the perturbed event horizon as a function of the circular-orbit frequency $\Omega$ of the binary. To do so, we simply need to invert the expression for $\Omega = \bar{\omega} + \calD \omega(\bar{\omega}) + \mathcal{O}(q^2)$, as given by Eq.~\eqref{Domega2} with $\chi \!=\! 4\mu\bar{\omega} [1-4(\mu\bar{\omega})^2]^{1/2}$ and $v^3 \!=\! \mu\bar{\omega} [1-4(\mu\bar{\omega})^2]^{-3/2} + \mathcal{O}(q)$. We readily find
\beq\label{omegaH}
	\bar{\omega} = \Omega \left\{ 1 - q \, u \left( 1 + 2u \right) \sqrt{\frac{1-4u^3}{1+u}} + \mathcal{O}(q^2) \right\} ,
\eeq
where we recall that $u = (\mu\Omega)^{2/3}$, with $\mu = \mu_2$ the irreducible mass of the background Kerr black hole.\footnote{Beware that in Sec.~\ref{subsec:kappa} and \ref{subsec:kappa_moon}, the symbol $\mu$ denotes the irreducible mass of the background Kerr black hole, i.e. $\mu = \mu_2$, and not the total irreducible mass like everywhere else.} Importantly, the 3PN expansion of this perturbative result is in full agreement with the large mass-ratio limit of the 3PN result \eqref{omega_a_PN}, namely
\beq
	\bar{\omega} = \omega_2 = \Omega \left\{ 1 - q \, \biggl( u + \frac{3}{2} u^2 + \frac{27}{8} u^3 \biggr) + \mathcal{O}(q^2,u^4) \right\} .
\eeq
This provides significant support to both the post-Newtonian and the black hole perturbative calculations. Substituting for \eqref{omegaH} into Eqs.~\eqref{kappa_Kerr} and \eqref{Dkappa2}, and expanding to first order in the mass ratio, we find
\beq\label{4mu2kappa2}
	4 \mu_2 \kappa_2 = \frac{1 - 8 u^3}{\sqrt{1 - 4 u^3}} - q_\mu \, \frac{u \left( 1 + 2 u + 4 u^2 - 8 u^3 - 16 u^4 \right)}{\sqrt{(1+u)(1-4u^3)}} + \mathcal{O}(q_\mu^2) \, .
\eeq
The expression for an isolated black hole, i.e. the first term in the right-hand side of Eq.~\eqref{4mu2kappa2}, is simply given by \eqref{4mukappa_Kerr} with $\omega = \bar{\omega} = \Omega$ at that order of approximation. Notice also that in the mass ratio $q$ we have substituted the Kerr black hole mass $M = m_2$ by the irreducible mass $\mu = \mu_2$  through
\beq\label{q}
	q = q_\mu \sqrt{1-4(\mu\bar{\omega})^2} = q_\mu \sqrt{1-4u^3} + \mathcal{O}(q_\mu^2) \, .
\eeq
As is clear from \eqref{4mu2kappa2}, the $\mathcal{O}(q_\mu)$ correction to the horizon surface gravity of the background Kerr black hole is negative. In the weak-field limit $u \to 0$, the perturbative result \eqref{4mu2kappa2} gives
\beq
	4 \mu_2 \kappa_2 = 1 - 6 u^3 - q_\mu \left( u + \frac{3}{2} u^2 + \frac{27}{8} u^3 - \frac{121}{16} u^4 - \frac{1005}{128} u^5 \right) + \mathcal{O}(u^6,q_\mu^2) \, ,
\eeq
in full agreement with the large mass-ratio limit \eqref{c2z2exp} of the 4PN result \eqref{c1z1}. This check of consistency gives additional support to the analogy \eqref{analogy} between point particle redshift and black hole horizon surface gravity.

\subsection{Surface gravity of the corotating moon}\label{subsec:kappa_moon}

We are now going to derive an explicit expression for the normalized surface gravity $4\mu_1\kappa_1$ of the orbiting moon (the lightest black hole), to linear order in the mass ratio. In Sec.~\ref{subsec:z-kappa} we mentionned a proof, given in Ref.~\cite{Po3.15}, that the normalized surface gravity $4\mu_1\kappa_1$ is in fact equal to the redshift $z \equiv z_1$ of a massive point particle undergoing a gravitational self-force effect, that follows a geodesic motion in a certain smooth effective metric. Hence, Eq.~\eqref{Adam} implies
\beq\label{z_exp}
	4\mu_1\kappa_1 = z = z_{(0)}(\Omega,\chi) + q \, z_{(1)}(\Omega,\chi) + \calO(q^2) \, ,
\eeq
where $z_{(0)} = [1 - 3 v^2 + 2 \chi v^3]^{1/2} / (1 + \chi v^3)$ is the redshift of a test mass on a circular equatorial orbit in a Kerr background of dimensionless spin $\chi$, while $q \, z_{(1)}(\Omega,\chi)$ is the (conservative) gravitational self-force correction at a given circular-orbit frequency $\Omega$. Substituting for the expressions $\chi = 4 \mu \bar{\omega} [1-4(\mu\bar{\omega})^2]^{1/2}$ and $v^3 = M\Omega /(1-\chi M\Omega) = \mu\Omega / ([1-4(\mu\bar{\omega})^2]^{1/2}- \chi \mu\Omega)$, and specifying to the corotating case by imposing Eq.~\eqref{omegaH}, we readily obtain
\beq\label{4mu1kappa1}
	4\mu_1\kappa_1 = (1-2u) \sqrt{(1+u)(1-4u^3)} + q \left\{ z_{(1)}(u) - 4 u^5 \frac{(1+2u)^2}{1+u} \right\} + \calO(q^2) \, .
\eeq

On the other hand, numerical data for $z_{(1)}(\Omega;\chi)$ has been computed in Refs.~\cite{Sh.al.12,vdMSh.15} for a sample of background Kerr black hole spins and circular-orbit frequencies. However, those data are too sparse for our purposes. Instead, we shall use the recent \emph{analytical} gravitational self-force results of \cite{Ka.al.16}, which provide the 8.5PN expansion of $z_{(1)}(\Omega;\chi)$ for any Kerr black hole spin $0 \leqslant \chi < 1$. Specifying to the corotating case, i.e., substituting $\chi = 4u^{3/2} [1-4u^3]^{1/2}$ in their Eqs.~(4.2)--(4.3), as well as $M = \mu [1-4u^3]^{-1/2}$, we find\footnote{Actually, the authors of Ref.~\cite{Ka.al.16} computed the 8.5PN expansion of the gravitational self-force contribution to the quantity $U \equiv 1/z$, which they denoted $\Delta U$. We simply have $z_{(1)} = - z_{(0)}^2 \Delta U$.}
\begin{align}\label{z_(1)}
	z_{(1)}&(u) = u - u^2 - u^3 + \biggl( \frac{52}{3} - \frac{41}{32} \pi^2 \biggr) u^4 + \biggl( - \frac{958}{15} + \frac{1291}{512} \pi^2 + \frac{128}{5} \gamma_\text{E} +\frac{64}{5} \ln{(16u)} \biggr) u^5 \nonumber \\ &+ \biggl( - \frac{2587787}{3150} + \frac{126779}{1536} \pi^2 - \frac{9976}{105} \gamma_\text{E} - \frac{23672}{15} \ln{2} + \frac{243}{7} \ln{3} - \frac{4988}{105} \ln{u} \biggr) u^6 + o(u^6) \, .
\end{align}
To avoid an exceedingly lengthy expression, we only display here the result up to 5PN order. The analytical values of the higher order coefficients are given in Eqs.~\eqref{z1}--\eqref{coeffs} of App.~\ref{app:coeffs}, up to 8.5PN order. In Sec.~\ref{sec:compare} below, we shall use this 8.5PN-accurate expansion to evaluate the normalized surface gravity \eqref{4mu1kappa1} of the smaller black hole.

Importantly, we can verify that the 4PN expansion of the perturbative result \eqref{4mu1kappa1}--\eqref{z_(1)}, together with Eq.~\eqref{q}, is in full agreement with the perturbative expansion \eqref{c1z1exp} of the 4PN result \eqref{c1z1}, namely
\begin{align}
	4\mu_1&\kappa_1 = 1 - \frac{3}{2} u - \frac{9}{8} u^2- \frac{27}{16} u^3 + \frac{363}{128} u^4 + \frac{603}{256} u^5 \nonumber \\ &+ q_\mu \, \biggl[ u - u^2 - u^3 + \biggl( \frac{46}{3} - \frac{41}{32} \pi^2 \biggr) u^4 + \biggl( - \frac{988}{15} + \frac{1291}{512} \pi^2 + \frac{128}{5} \gamma_\text{E} +\frac{64}{5} \ln{(16u)} \biggr) u^5 \biggr] \nonumber \\ & + o(u^5,q_\mu) \, .
\end{align}

\subsection{Rescaling of the perturbative results}\label{subsec:sym}

The 4PN formula \eqref{c1z1} is naturally expressed in terms of the PN parameter $y = (\mu \Omega)^{2/3}$, where the orbital frequency is normalized by the \emph{total} irreducible mass $\mu = \mu_1 + \mu_2$. Hence, to compare our perturbative formulas \eqref{4mu2kappa2} and \eqref{4mu1kappa1} to the 4PN and NR results, we first use the relation $u = y \, (1+q_\mu)^{-2/3}$, expand to linear order in the irreducible mass ratio $q_\mu$, and obtain
\begin{subequations}\label{4mukappatemp}
	\begin{align}
		4 \mu_1 \kappa_1 &= (1-2y) \sqrt{(1+y)(1-4y^3)} + \mathcal{O}(q_\mu^2) \nonumber \\ &\quad + q_\mu \biggl[ \frac{y \, (1 + 2 y + 4 y^2 - 8 y^3 - 16 y^4)}{\sqrt{(1 + y)(1 - 4 y^3)}} - 4 y^5 \sqrt{1-4y^3} \, \frac{(1+2y)^2}{1+y} + \sqrt{1-4y^3} \, z_{(1)}(y) \biggr] \, , \label{4mu1kappa1temp} \\
		4 \mu_2 \kappa_2 &= \frac{1 - 8 y^3}{\sqrt{1 - 4 y^3}} - q_\mu \biggl[ \frac{y \, (1 + 2 y + 4 y^2 - 8 y^3 - 16 y^4)}{\sqrt{(1 + y)(1 - 4 y^3)}} - \frac{4 y^3 \left( 3 - 8 y^3 \right)}{(1 - 4 y^3)^{3/2}} \biggr] + \mathcal{O}(q_\mu^2) \, . \label{4mu2kappa2temp}
	\end{align}
\end{subequations}
Remarkably, we observe that the first terms in the $\calO(q_\mu)$ contributions therein are identical. While the former comes from the substitution $u \to y$ in the $\calO(q_\mu^0)$ contribution to \eqref{4mu1kappa1temp}, the latter comes from combining the same substitution in the $\calO(q_\mu^0)$ contribution to \eqref{4mu2kappa2temp} with the $\calO(q)$ correction to the horizon angular velocity \eqref{omegaH}.

\hspace{-0.15cm}Motivated by a large body of work \cite{FiDe.84,An.al.95,Fa.al.04,Sp.al2.11,Le.al.11,Le.al2.12,Le.al.13,Na.13,vdM.17} suggesting that the scaling $q_\mu \to \eta = q_\mu/(1+q_\mu)^2$ considerably extends the domain of validity of black hole perturbative calculations, we wish to restore in the expansions \eqref{4mukappatemp} some of the missing information regarding the symmetry property by exchange $1 \leftrightarrow 2$ of the two bodies. As argued in App.~\ref{app:E-J}, the general form \eqref{flap} of the renormalized redshifts $c_a z_a$, together with the analogy \eqref{analogy}, suggest that the surface gravity of each black hole should naturally be written in the form
\beq\label{4mukappaexp}
	4\mu_a\kappa_a = a(y) \pm b(y) \, \Gamma + c(y) \, \eta \pm d(y) \, \Gamma \eta + \calO(\eta^2) \, .
\eeq
The key point is that the a priori unknown functions $a(y)$, $b(y)$, $c(y)$ and $d(y)$ can be \emph{uniquely} determined by expanding Eqs.~\eqref{4mukappaexp} up to linear in the mass ratio, using $\Gamma = 1 - 2q_\mu + \calO(q_\mu^2)$, and equating the resulting expressions to the formulas \eqref{4mukappatemp}. Doing so, we readily obtain
\begin{subequations}\label{a-d}
	\begin{align}
		a(y) &= \frac{1}{2} (1-2y) \sqrt{(1+y)(1-4y^3)} + \frac{1-8y^3}{2\sqrt{1-4y^3}} \, , \label{coeffa} \\
		b(y) &= \frac{1}{2} (1-2y) \sqrt{(1+y)(1-4y^3)} - \frac{1-8y^3}{2\sqrt{1-4y^3}} \, , \label{coeffb} \\
		c(y) &= \sqrt{1-4y^3} \, \biggl\{ \frac{z_{(1)}(y)}{2} - 2 y^5 \, \frac{(1+2y)^2}{1+y} \biggr\} + \frac{2y^3(3-8y^3)}{{(1-4y^3)}^{3/2}} \, , \label{coeffc} \\
		d(y) &= \sqrt{1-4y^3} \, \biggl\{ \frac{z_{(1)}(y)}{2} - 2 y^5 \, \frac{(1+2y)^2}{1+y} \biggr\} + \frac{1-4y^4-8y^5}{\sqrt{(1+y)(1-4y^3)}} - \frac{1-6y^3+16y^6}{{(1-4y^3)}^{3/2}} \, .
	\end{align}
\end{subequations}
In particular, the coefficients $a$ and $b$ are simply given by the half-sum and the half-difference of the $\calO(q_\mu^0)$ contributions to Eq.~\eqref{4mukappatemp}, respectively, while the coefficient $c$ is given by the half-sum of the $\calO(q_\mu)$ contributions. Expanding \eqref{4mukappaexp}--\eqref{a-d} in powers of $q_\mu$, one recovers the original expressions \eqref{4mukappatemp}, by construction. One the other hand, expanding in powers of $y$, one recovers the 4PN formula \eqref{c1z1}, up to uncontrolled terms $\calO(\eta^2)$ and higher. For equal-mass binaries, $\Gamma = 0$ and the expressions \eqref{4mukappaexp}--\eqref{a-d} for $4 \mu_1 \kappa_1$ and $4 \mu_2 \kappa_2$ coincide.

\section{Comparisons}
\label{sec:compare}

In this section, we shall compare the predictions from our numerical relativity initial data, the post-Newtonian approximation, and black hole perturbation theory for corotating black hole binaries. In Sec.~\ref{subsec:kappas} we compare the computed surface gravities for irreducible mass ratios $q_\mu \in \{1,1/2,1/10\}$. We then revisit the problem of diagnostics of spin measurements in quasi-equilibrium initial data of corotating black hole binaries in Sec.~\ref{subsec:spin}. Throughout this section, we use the total irreducible mass $\mu = \mu_1 + \mu_2$ to adimensionalize all frequencies.

\subsection{Surface gravity}
\label{subsec:kappas}

Motivated by the analogy pointed out in Refs.~\cite{Le.al.12,Bl.al.13} between black hole surface gravity and point particle redshift, the authors of Ref.~\cite{Zi.al.16} introduced a concept of ``redshift'' for a dynamical black hole, from a properly normalized and averaged notion of surface gravity, in the context of NR simulations of nonspinning black hole binaries. They compared their numerical results for these ``black hole redshifts'' to the known 3PN prediction \cite{Bl.al.10} for the redshifts of point particle binaries for circular orbits, and found excellent agreement during the entire inspiral, with relative differences $\calO(10^{-3})$.

In this work, we adopt a strategy that, in a sense, is the opposite to that used in Ref.~\cite{Zi.al.16}. In Sec.~\ref{sec:NR} we have computed the surface gravity $4\mu_a \kappa_a$ of each corotating black hole. Then, motivated by the analogy \eqref{analogy}, we are going to compare those numerical results to the 4PN-accurate prediction \eqref{c1z1} for the redshift $c_a z_a$ of each corotating point particle. Moreover, we will compare our nonlinear result for the surface gravity of each black hole to the predictions \eqref{4mukappatemp} and \eqref{4mukappaexp}--\eqref{a-d} from linear black hole perturbation theory. 

\subsubsection{Mass ratio 1:1}

\begin{figure}[t!]
	\includegraphics[width=0.82\linewidth]{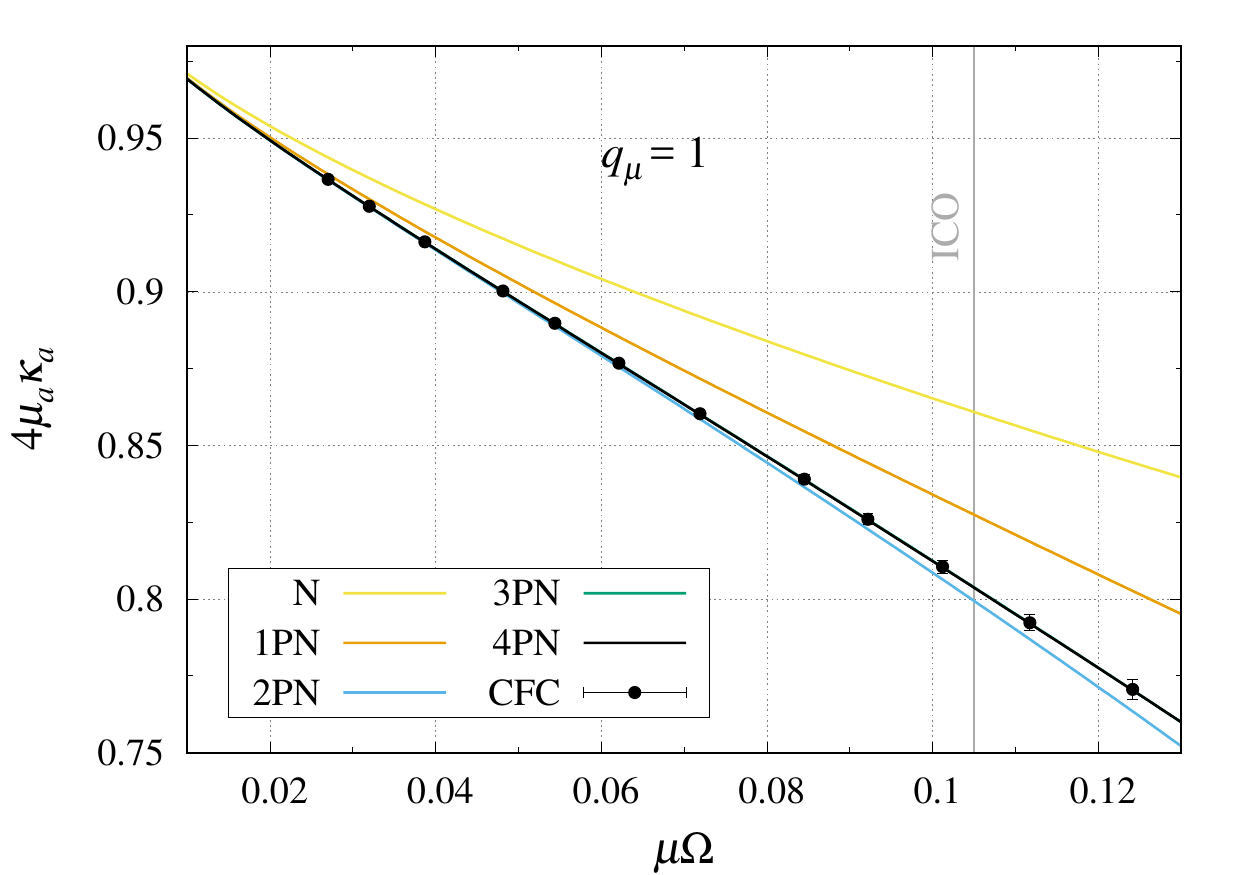}
	\caption{Normalized horizon surface gravity $4\mu_a\kappa_a$ in equal-mass corotating binaries, as a function of the circular-orbit frequency $\mu\Omega$, computed using the CFC and the PN approximations. The data points with error bars display $4\mu_a\av{\kappa_a} [1 \pm \bar\sigma(\kappa_a)]$. The various colored curves show the successive PN approximations to $c_a z_a$, up to 4PN order, computed from Eq.~\eqref{c1z1} with $\eta = \frac{1}{4}$ and $\Gamma = 0$. The vertical gray line displays the location of the innermost circular orbit, at $\mu\Omega_\text{ICO} \simeq 0.105$.}
	\label{fig:kappa_q1}
\end{figure}

We start with the simplest case of equal-mass binaries, for which $\kappa_1 = \kappa_2$ by symmetry. Figure \ref{fig:kappa_q1} shows the normalized horizon surface gravity $4\mu_a \kappa_a$ of both black holes as a function of the orbital frequency $\mu\Omega$, as computed in Sec.~\ref{sec:NR} using the CFC approximation, and in Sec.~\ref{sec:PN} using the PN approximation, up to 4PN order. More precisely, the data points with error bars display $4\mu_a\av{\kappa_a} [1 \pm \bar\sigma(\kappa_a)]$, and the various colored curves show the successive PN approximations to $c_a z_a$, up to 4PN order, computed from \eqref{c1z1} with $\eta = \frac{1}{4}$ and $\Gamma = 0$. The vertical gray line shows the location of the ICO for this particular configuration, at $\mu\Omega_{\text{ICO}} \simeq 0.105$, as computed from the binding energy $E(\Omega)$ of the sequence of quasi-equilibrium initial data (see Tab.~\ref{tab:ICO}).

The NR data shows that the surface gravity decreases fairly steeply with increasing orbital frequency, as the effect of the tidal field of the companion gets stronger. In the Newtonian limit where $\mu\Omega \to 0$, we recover the expected behavior $4\mu_a\kappa_a \to 1$, as appropriate for isolated nonspinning black holes. While the Newtonian and 1PN approximations to $4\mu_a\kappa_a$ compare quite poorly to the NR results, except at low frequencies (i.e., at large orbital separations), the 2PN, 3PN and 4PN approximations perform very well, even for highly relativistic orbits beyond the ICO. In particular, the 3PN and 4PN predictions---which cannot be distinguished on the graph---are both within the NR error bars over the entire orbital-frequency range for which we have data.

\subsubsection{Mass ratios 2:1 and 10:1}

\begin{figure}[t!]
	\includegraphics[width=0.82\linewidth]{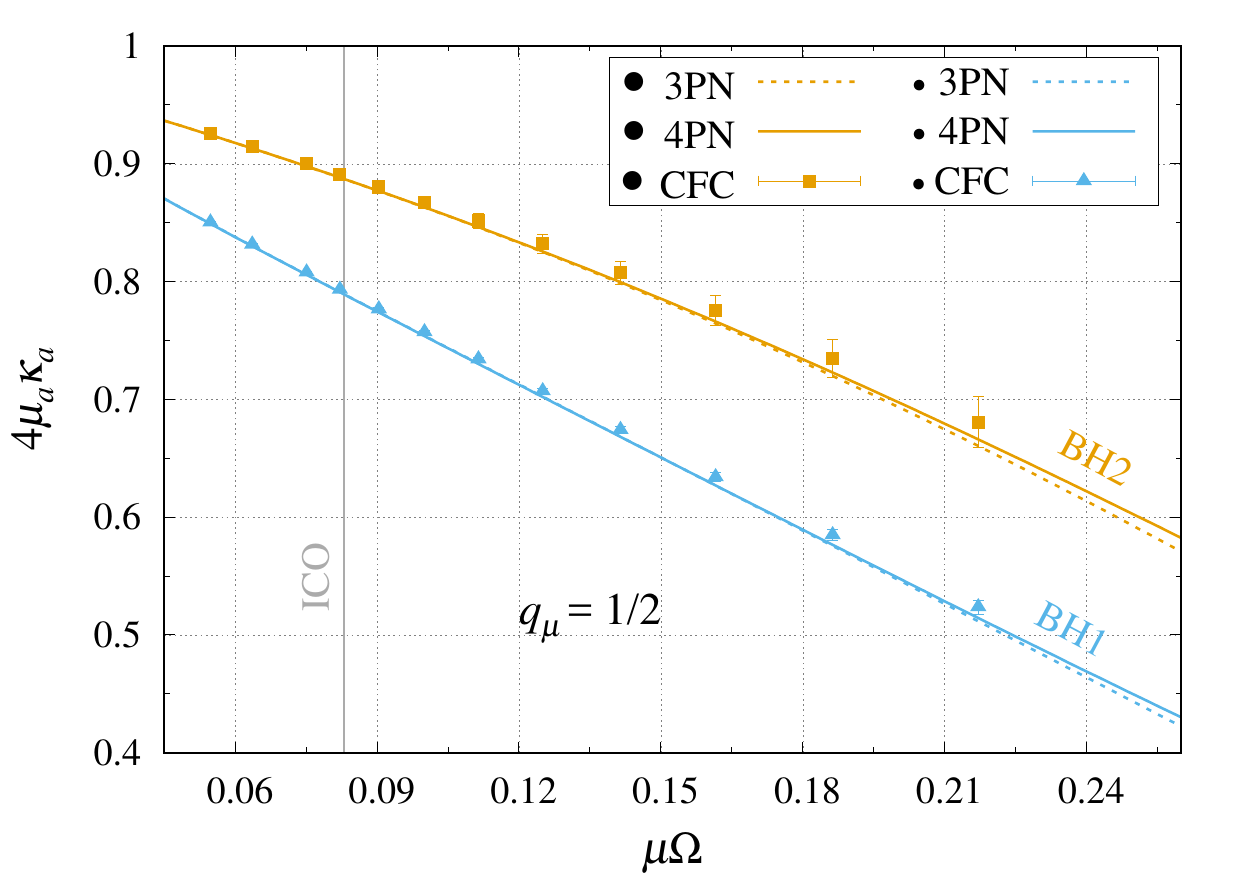}
	\caption{Same as in Fig.~\ref{fig:kappa_q1} but for a mass ratio $2:1$, for which $\mu\Omega_\text{ICO} \simeq 0.083$. The black holes can be distinguished by the size of the filled black circles in the legend. By convention $\mu_1 \leqslant \mu_2$ so black hole 1 (BH1) is lighter than black hole 2 (BH2). For all but the rightmost pale-blue data point the CFC error bars are smaller than the size of the point symbol.} 
	\label{fig:kappa_q2}
	\vspace{-0.5cm}
\end{figure}

Next, we consider the case of a binary with mass ratio $q_\mu = 1/2$, for which $\mu\Omega_\text{ICO} \simeq 0.083$. Figure \ref{fig:kappa_q2} displays the NR results for each black hole, which can be distinguished by the size of the filled black circle in the figure's legend. First, we notice that the (normalized) surface gravity of the heaviest black hole is larger than that of the lightest black hole. This indicates that the tidal field of the largest body has a bigger effect on the smaller body than the other way around, in agreement with physical intuition. Figure \ref{fig:kappa_q2} also shows the 3PN and 4PN predictions for each black hole. We observe a slight improvement from 3PN to 4PN order, for both black holes. Overall, the agreement with the NR results is very good, all the way to the ICO, and even beyond. The relative differences reach $2\%$ for large orbital frequencies, say at $\mu \Omega \sim 0.2 \sim 2.4 \, \mu\Omega_\text{ICO}$, corresponding to an orbital separation $r_\Omega \equiv \mu / (\mu\Omega)^{2/3} \sim 3 \mu$.

\begin{figure}[h!]
	\includegraphics[width=0.82\linewidth]{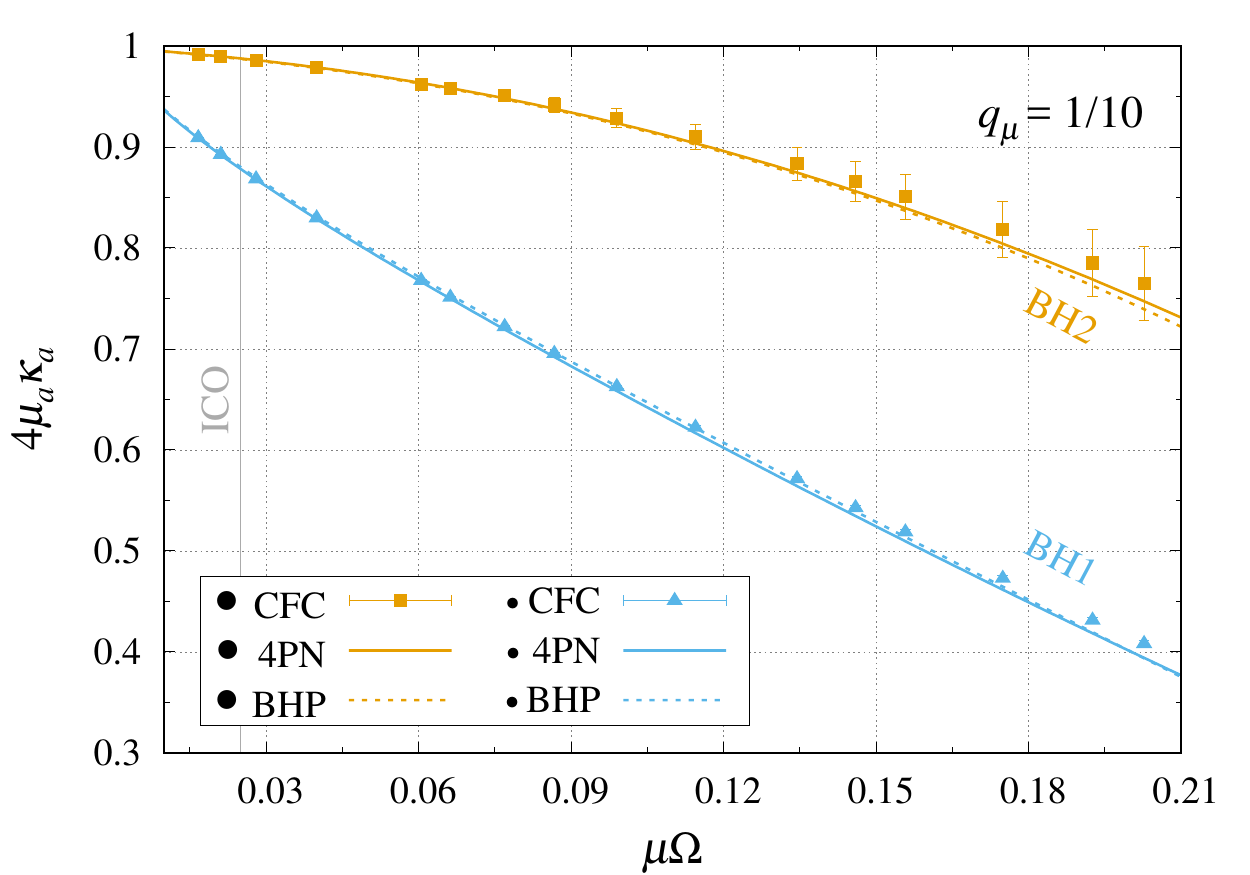}
	\caption{Same as in Fig \ref{fig:kappa_q2} but for a mass ratio $10:1$, for which $\mu\Omega_\text{ICO} \simeq 0.025$. The dashed curves labelled ``BHP'' show the predictions from linear black hole perturbation theory, as given in \eqref{4mukappatemp}. For the lighter black hole (BH1), the error bars on the CFC data points are all smaller than the size of the point symbol.}
	\label{fig:kappa_q10}
	\vspace{-1.5cm}
\end{figure}

Our third and last case corresponds to a mass ratio $q_\mu = 1/10$, for which $\mu\Omega_\text{ICO} \simeq 0.025$. Figure \ref{fig:kappa_q10} shows the NR results, as well as the analytic predictions from the PN approximation and from linear black hole perturbation theory. Just like for the cases $q_\mu = 1$ and $1/2$, the 4PN predictions perform very well over the full range of circular-orbit frequencies considered. More remarkably, the predictions from \emph{linear} black hole perturbation theory are in excellent agreement with the NR data, despite the moderate mass ratio, all the way to separations $r_\Omega \sim 3 \mu$ (i.e. well inside the ICO), for both black holes. In fact, the perturbative prediction for the smaller body is in better agreement with the NR data than the 4PN prediction.


\subsubsection{Mass ratio rescaling}

\begin{figure}[h!]
	\includegraphics[width=0.82\linewidth]{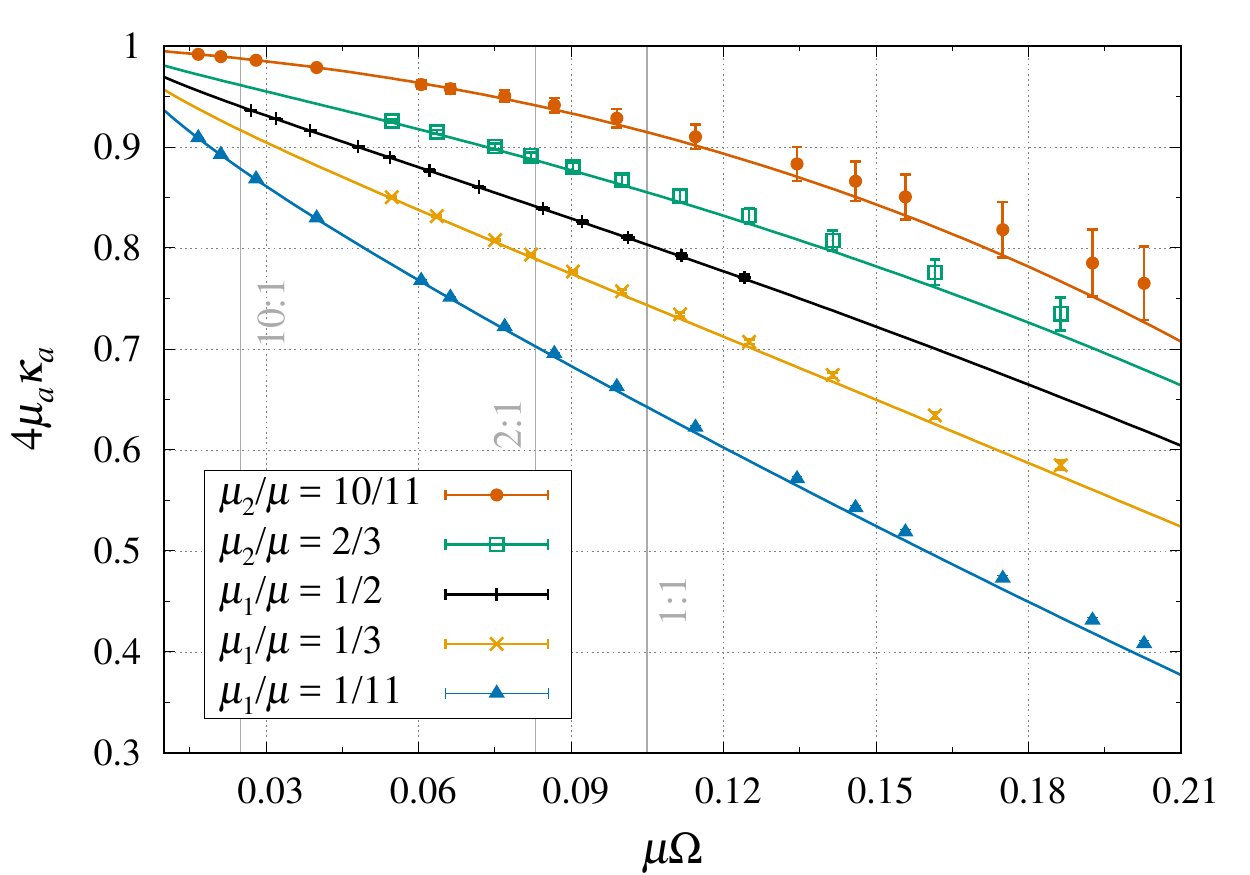}
	\caption{Same as in Figs.~\ref{fig:kappa_q1}, \ref{fig:kappa_q2} and \ref{fig:kappa_q10}, all binary configurations combined. The solid curves show the predictions from linear black hole perturbation theory, following the mass-ratio rescaling discussed in Sec.~\ref{subsec:sym}, as given in Eqs.~\eqref{4mukappaexp}--\eqref{a-d}. The vertical gray lines show the location of the innermost circular orbit for each irreducible mass ratio $\mu_2/\mu_1$.} 
	\label{fig:kappa_multi_q}
\end{figure}

Finally, in Fig.~\ref{fig:kappa_multi_q} we collect all the NR results discussed previously, together with the predictions from black hole perturbation theory in each case, following the mass-ratio rescaling introduced in Sec.~\ref{subsec:sym}. Remarkably, the rescaled perturbative predictions \eqref{4mukappaexp}--\eqref{a-d} are in excellent agreement with the NR data in \emph{all} cases, for both black holes. The agreement in the equal-mass case (black curve and symbols) is especially stricking. Indeed, the rescaled \emph{linear} perturbative predictions are compatible with the nonlinear results (agreement within the NR error bars) over most of the frequency range. The largest disagreement occures for the smallest black hole, for a mass ratio $q_\mu = 1/10$, for which the relative difference grows up to $\sim 3\%$ at $r_\Omega \sim 3 \mu$. Restricting ourselves to physically relevant orbits outside the ICO, the agreement reaches the $0.3\%$ level or better for all mass ratios considered.

This illustrates, once again, that the domain of validity of black hole perturbative calculations appears to extend well beyond the extreme mass-ratio limit. As argued in Ref.~\cite{Le2.14}, this opens the exciting prospect of using black hole perturbation theory and the gravitational self-force framework to model the gravitational-wave emission from intermediate mass-ratio inspirals, or even compact binaries with comparable masses.

This observation is also supported by recent works showing that (i) the Einstein equation is, in a sense, not so nonlinear \cite{Ha2.14} and that (ii) the relativistic gravitational dynamics of an arbitrary-mass-ratio two-spinning-black-hole system is equivalent, at first post-Minkowskian order, to that of a spinning test black hole in a background Kerr spacetime \cite{Da.16,Vi.18}.

\begin{figure}[h!]
	\includegraphics[width=0.8\linewidth]{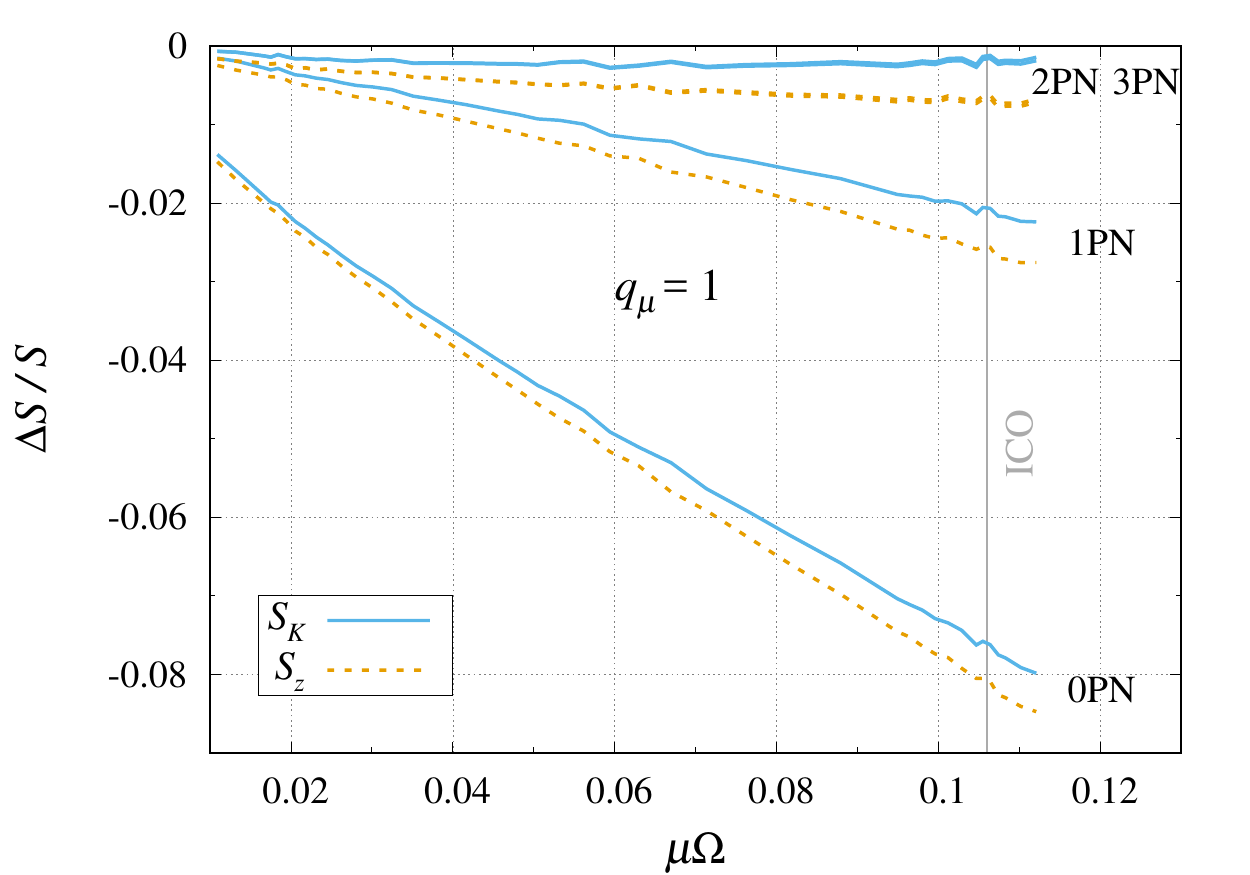}
	\caption{Relative differences $\Delta S / S$ between several quasi-local measurements of spins for equal-mass corotating black hole binaries, as functions of the circular-orbit frequency $\mu\Omega$. The 2PN and 3PN curves are almost on top of each other.} 
	\label{fig:spins}
	\vspace{-0.25cm}
\end{figure}

\subsection{Quasi-local spin}
\label{subsec:spin}

In binary black hole spacetimes, there is no canonical definition of spin for each black hole. However, several \textit{quasi-local} notions of spins have been introduced in the litterature \cite{BrYo.93,As.al.01,CoWh.07} (see \cite{Sz.04,AsKr.04} for reviews). In particular, the authors of Ref.~\cite{Ca.al2.06} used two methods to estimate the spins of equal-mass, corotating black holes binaries: one based on an approximate axial Killing vector, $S_K$, and one based on a flat space Killing vector constructed from Cartesian-like coordinates, $S_z$. In the case of a sequence of quasi-equilibrium initial data for equal-mass, corotating black hole binaries, those spin diagnostics are listed in Table IV of Ref.~\cite{Ca.al2.06}, for a particular choice of lapse boundary condition on the excision surfaces. The relative difference between $S_K$ and $S_z$ remains below $0.6\%$ over the orbital frequency range $0.01 \lesssim \mu\Omega \lesssim 0.11$.

On the other hand, the discussion in Sec.~\ref{subsec:2PPcor} suggests that each black hole in a corotating binary can, to some extent, be modelled as an isolated hole immersed in a tidal environment. Then, a third natural spin measurement is provided by the Kerr-inspired formula \eqref{Sa}, in which the proper rotation frequency $\omega_a$ must be expressed as a function of the circular-orbit frequency $\Omega$ of the binary. Such a relationship was precisely determined, up to 3PN order, in Eq.~\eqref{omega_a_PN}. We denote by $S$ the spin measurement obtained by inserting Eq.~\eqref{omega_a_PN} into the formula \eqref{Sa}, without performing any subsequent PN expansion.

Figure \ref{fig:spins} shows the relative differences $\Delta S/S \equiv S_K/S-1$ and $S_z/S-1$ while using the Newtonian (labelled 0PN), 1PN, 2PN and 3PN approximations for the black hole's proper rotation frequencies \eqref{omega_a_PN} in the expression \eqref{Sa} for the spin $S$ of a black hole of irreducible mass $\mu_a$ and proper rotation frequency $\omega_a$. This figure updates Fig.~4 of Ref.~\cite{Ca.al2.06}. We make the three following observations:
\begin{itemize}
	\item At a given PN order, the $S_K$ spin diagnostic is always in better agreement with the $S$ spin diagnostic than the $S_z$ spin diagnostic;
	\vspace{-0.04cm}
	\item Increasing the PN order systematically reduces the disagreements between the multiple spin diagnostics, from up to $8\%$ at Newtonian order to less than $0.3\%$ at 3PN order;
	\vspace{-0.04cm}
	\item Given the small numerical value of the 3PN coefficient for equal masses in Eq.~\eqref{omega_a_PN}, the difference between the 2PN and 3PN-accurate predictions for the $S$ spin diagnostic is small over the orbital frequency range considered, such that the 2PN and 3PN curves are almost on top of each other.
\end{itemize}

In particular, this suggests that the corotation condition \eqref{cond_cor} for spinning point masses captures most of the physics of corotating black hole binaries, which may (in part) explain the remarkable agreement between the NR and PN results for the horizon surface gravity, as discussed in Sec.~\ref{subsec:kappas}. That being said, despite a trend that shows better agreement at increasingly higher PN orders, one should not expect such comparisons to reach an arbitrarily high level of agreement, because each of those spins diagnostics comes with caveats.

\acknowledgements
ALT acknowledges financial support through a Marie Curie FP7 Integration Grant within the 7th European Union Framework Programme (PCIG13-GA-2013-630210).

\appendix

\section{Energy and angular momentum of corotating binaries}
\label{app:E-J}

In this appendix, we give the expressions for the binding energy $E$ and angular momentum $J$ of corotating compact binaries, as functions of the circular-orbit frequency $\Omega$, up to 4PN order. First, recall that for nonspinning binaries, the 4PN-accurate expressions of the binding energy and angular momentum are known and read \cite{Da.al.14,Be.al.17}
\begin{subequations}\label{E-J_nonspin}
	\begin{align}	
		\!\!\!E^\text{ns} &= -\frac{\mu \eta \, y}{2} \biggl\{ 1 + \left( - \frac{3}{4} - \frac{\eta}{12} \right) y + \left( - \frac{27}{8} + \frac{19}{8} \eta - \frac{\eta^2}{24} \right) y^2 \nonumber \\ &\qquad\qquad\quad\,\; + \left( - \frac{675}{64} + \biggl[ \frac{34445}{576} - \frac{205}{96} \pi^2 \biggr] \eta - \frac{155}{96} \eta^2 - \frac{35}{5184} \eta^3 \right) y^3 \nonumber \\ &\qquad\qquad\quad\,\; + \left( - \frac{3969}{128} + \left[ - \frac{123671}{5760} + \frac{9037}{1536} \pi^2 + \frac{896}{15} \gamma_\text{E} + \frac{448}{15} \ln(16 y) \right] \eta \right. \nonumber \\ &\left.\qquad\qquad\qquad\quad + \left[ - \frac{498449}{3456} + \frac{3157}{576} \pi^2 \right] \eta^2 + \frac{301}{1728} \eta^3 + \frac{77}{31104} \eta^4 \right) y^4 + o(y^4)\biggr\} \, , \\
		\!\!\!\Omega J^\text{ns} &= \mu \eta \, y \, \biggl\{ 1 + \left( \frac{3}{2} + \frac{\eta}{6} \right) y + \left( \frac{27}{8} - \frac{19}{8} \eta + \frac{\eta^2}{24} \right) y^2 \nonumber \\ &\qquad\qquad\;\; + \left( \frac{135}{16} + \biggl[ - \frac{6889}{144} + \frac{41}{24} \pi^2 \biggr] \eta + \frac{31}{24} \eta^2 + \frac{7}{1296} \eta^3 \right) y^3 \nonumber \\ &\qquad\qquad\;\; + \left( \frac{2835}{128} + \left[ \frac{98869}{5760} - \frac{6455}{1536} \pi^2 - \frac{128}{3} \gamma_\text{E} - \frac{64}{3} \ln(16 y) \right] \eta \right. \nonumber \\ &\left.\qquad\qquad\qquad\; + \left[ \frac{356035}{3456} - \frac{2255}{576} \pi^2 \right] \eta^2 - \frac{215}{1728} \eta^3 - \frac{55}{31104} \eta^4 \right) y^4 + o(y^4)\biggr\} \, ,
	\end{align}
\end{subequations}
where we recall that $y = (\mu \Omega)^{2/3}$, with $\mu = \mu_1 + \mu_2$ the total irreducible mass and $\eta \!=\! \mu_1 \mu_2 / \mu^2$. Since for nonspinning binaries $m_a = \mu_a$, the sets of variables $(\mu,\eta,y)$ and $(m,\nu,x)$ coincide. Notice the logarithmic running at 4PN order, related to the occurence of gravitational-wave tails at that order. 

On the other hand, by combining the leading order and next-to-leading order spin-orbit contributions to the binding energy and angular momentum with the contributions coming from the replacement of the masses $m_a$ by the irreducible masses $\mu_a$ and the spins $\chi_a$ by their expressions as functions of the orbital frequency $\Omega$ (see Ref.~\cite{Bl.al.13}), we obtain the \textit{additional} spin-related contributions $E^\text{cor}$ and $J^\text{cor}$ in the corotating case. Those explicitly read
\begin{subequations}\label{E-J_corot}
	\begin{align}
		\!\!\!E^\text{cor} &= \mu \left( 2 - 6 \eta \right) y^3 + \mu \eta \left( - 10 + 25 \eta \right) y^4 + \mu \eta \left( - 21 + \frac{320}{3} \eta - \frac{325}{6} \eta^2 \right) y^5 + o(y^5) , \\ 
		\!\!\!\Omega J^\text{cor} &= \mu \left( 4 - 12 \eta \right) y^3 + \mu \eta \left( - 16 + 40 \eta \right) y^4 + \mu \eta \left( - 30 +\frac{224}{3} \eta - \frac{455}{12} \eta^2 \right) y^5 + o(y^5) .
	\end{align}
\end{subequations}
These 2PN, 3PN and 4PN spin-related contributions have to be added to the 4PN-accurate expressions \eqref{E-J_nonspin} for spinless binaries, such that for corotating binaries the binding energy and angular momentum are given by $E = E^\text{ns} + E^\text{cor}$ and $J = J^\text{ns} + J^\text{cor}$. The 3PN results were first (incorrectly) computed in Ref.~\cite{Bl.02}, by using the leading-order formula $\omega_a = \Omega$ for the proper rotation frequencies of the spinning particles, and were later corrected in Ref.~\cite{Bl.al.13} using Eq.~\eqref{omega_a_PN}. Here we extended these calculations to 4PN order.

One can then check that the formulas \eqref{E-J_nonspin} and \eqref{E-J_corot}, together with the expressions \eqref{c1z1} for the particles' (renormalized) redshifts, obey the first law \eqref{1st_law_2PP_cor} for corotating binaries, which is equivalent to the partial differential equations
\begin{subequations}\label{PDEs}
	\begin{align}
		\frac{\partial E}{\partial \Omega} \bigg|_{\mu_a} &= \Omega \, \frac{\partial J}{\partial \Omega} \bigg|_{\mu_a} \, , \\
		\frac{\partial E}{\partial \mu_a} \bigg|_\Omega &= \Omega \, \frac{\partial J}{\partial \mu_a} \bigg|_\Omega + c_a z_a - 1 \, .
	\end{align}
\end{subequations}
Similarly, one can easily check that the first integral \eqref{1st_int_2PP_cor} is also satisfied.

The partial differential equations \eqref{PDEs} can also be combined to establish that the redshifts $c_a z_a$ take a simple form when expressed in terms of the variables $y$ and $\eta$. Indeed, performing the change of variables $(\Omega,\mu_1,\mu_2) \to (y,\mu,\eta)$, we have (see also Sec.~IV B in Ref.~\cite{Le.al.12})
\beq\label{flip}
	c_a z_a - 1 = \frac{\partial E}{\partial \mu_a} - \Omega \, \frac{\partial J}{\partial \mu_a} =  \hat{\mathcal{E}} + \frac{2y}{3} \, \frac{\partial \hat{\mathcal{E}}}{\partial y} + \frac{1}{2} (1\pm\Gamma-4\eta) \, \frac{\partial \hat{\mathcal{E}}}{\partial \eta} \, ,
\eeq
where we introduced the convenient combination $\mathcal{\hat{E}} \equiv (E - \Omega J) / \mu$, heuristically the binary's reduced binding energy in a corotating frame. Now, on dimensional grounds alone, we know that $\mathcal{\hat{E}}(y,\eta) = \sum_{k \geqslant 0} \eta^k e_{(k)}(y)$, with $e_{(k)}$ dimensionless functions of the dimensionless variable $y$. This can be checked explicitly, for instance up to 4PN order, using the expressions \eqref{E-J_nonspin} and \eqref{E-J_corot}. Hence, Eq.~\eqref{flip} implies for the renormalized redshifts
\beq\label{flap}
	c_a z_a = f(y,\eta) \pm \Gamma \, g(y,\eta) \, ,
\eeq
for some functions $f$ and $g$ that can (at least formally) be written as $f(y,\eta) = \sum_{k \geqslant 0} \eta^k f_{(k)}(y)$ and $g(y,\eta) \!=\! \sum_{k \geqslant 0} \eta^k g_{(k)}(y)$. This can also be checked explicitly, up to 4PN order, from the formula \eqref{c1z1}. Recalling the analogy \eqref{analogy}, or following a similar argument starting from the first law \eqref{1st_law_2BH} for corotating black hole binaries, the general result \eqref{flap} motivated us in Sec.~\ref{subsec:sym} to rewrite the perturbative expansions \eqref{4mukappatemp} in the form \eqref{4mukappaexp}--\eqref{a-d}.

\section{Redshift of a corotating point particle in a Kerr background}
\label{app:coeffs}

In this appendix, we consider the PN expansion of the gravitational self-force contribution $z_{(1)}(\Omega)$ to the redshift of a point particle on the unique circular equatorial corotating orbit around a Kerr black hole of given spin. To do so, we use the results of \cite{Ka.al.16}, who computed the 8.5PN expansion of the quantity $\Delta U = - z_{(1)} / z_{(0)}^2$, where $z_{(0)}(\Omega)$ is given in the first line below \eqref{z_exp}, for a generic circular equatorial orbit, i.e., for any value of the Kerr black hole spin. Here, we specify to the corotating case by substituting the formula $\chi = 4u^{3/2} [1-4u^3]^{1/2}$ into their Eqs.~(4.1)--(4.2). Moreover, since we normalize the orbital frequency $\Omega$ using the irreducible mass $\mu$, rather than the black hole mass $M$, we also substitute $M = \mu [1-4u^3]^{-1/2}$. Up to 8.5PN order, we readily obtain
\beq\label{z1}
	z_{(1)}(u) = \sum_{k=1}^{18/2} \left( a_k + b_k \ln{u} + c_k \ln^2{\!u} \right) u^k + o(u^{18/2}) \, , 
\eeq
where the index $k$ runs over integer \textit{and} half-integer values, with the non-zero coefficients
\begin{subequations}\label{coeffs}
	\begin{align}
		a_1 &= - a_2 = - a_3 = 1 \, , \\
		a_4 &= \frac{52}{3} - \frac{41}{32} \pi^2 \, , \\
		a_5 &= - \frac{958}{15} + \frac{1291}{512} \pi^2 + \frac{128}{5} \gamma_\text{E} + \frac{256}{5} \ln{2} \, , \\
		a_6 &= - \frac{2587787}{3150} + \frac{126779}{1536} \pi^2 - \frac{9976}{105} \gamma_\text{E} - \frac{23672}{105} \ln{2} + \frac{243}{7} \ln{3} \, , \\
		a_{6.5} &= \frac{13696}{525} \pi \, , \\
		a_7 &= - \frac{39227674}{14175} + \frac{810460367}{1769472} \pi^2 - \frac{2800873}{262144} \pi^4 - \frac{357688}{2835} \gamma_\text{E} \nonumber \\ &\quad\, + \frac{6808}{81} \ln{2} - \frac{1944}{7} \ln{3} \, , \\
		a_{7.5} &= - \frac{368693}{3675} \pi \, , \\
		a_8 &= - \frac{4792193805194}{191008125} + \frac{4809089357497}{2477260800} \pi^2 + \frac{561618641}{16777216} \pi^4 \nonumber \\ &\quad\, + \frac{12682074238}{5457375} \gamma_\text{E} - \frac{109568}{525} \gamma_\text{E}^2 - \frac{438272}{525} \gamma_\text{E} \ln{2} \nonumber \\ &\quad\, + \frac{17972176126}{5457375} \ln{2} - \frac{438272}{525} \ln^2{2} + \frac{15704361}{24640} \ln{3} \nonumber \\ &\quad\, + \frac{1953125}{19008} \ln{5} + \frac{2048}{5} \zeta(3) \, , \\
		a_{8.5} &= - \frac{361209292}{3274425} \pi \, , \\
		a_9 &= \frac{3591487302848083}{17381739375} + \frac{1242000835942981}{92484403200} \pi^2 - \frac{22787284128473}{6442450944} \pi^4 \nonumber \\ &\quad\, - \frac{1857416227468}{496621125} \gamma_\text{E} + \frac{7443104}{11025} \gamma_\text{E}^2 + \frac{2555584}{735} \gamma_\text{E} \ln{2} - \frac{37908}{49} \gamma_\text{E} \ln{3} \nonumber \\ &\quad\, - \frac{4281723492044}{496621125} \ln{2} + \frac{6087776}{1575} \ln^2{2} - \frac{37908}{49} \ln{2} \ln{3} - \frac{18954}{49} \ln^2{3} \nonumber \\ &\quad\, + \frac{977811113907}{196196000} \ln{3} - \frac{513671875}{370656} \ln{5} - \frac{87616}{105} \zeta(3) \, ,  \\
		a_{9.5} &= - \frac{2560100455176683}{1048863816000} \pi - \frac{23447552}{55125} \pi \gamma_\text{E} - \frac{46895104}{55125} \pi \ln{2} + \frac{219136}{1575} \pi^3 \, , \\
 		b_5 &= \frac{64}{5} \, , \quad b_6 = - \frac{4988}{105} \, , \quad b_7 = - \frac{178844}{2835} \, , \\
 		b_8 &= \frac{6341037119}{5457375} - \frac{109568}{525} \gamma_\text{E} - \frac{219136}{525} \ln{2} \, , \\
		b_9 &= - \frac{935064864134}{496621125} + \frac{7443104}{11025} \gamma_\text{E} + \frac{1277792}{735} \ln{2} - \frac{18954}{49} \ln{3} \, , \\		b_{9.5} &= - \frac{11723776}{55125} \pi \, , \\
 		c_8 &= - \frac{27392}{525} \, , \quad c_9 = \frac{1860776}{11025} \, ,
	\end{align}
\end{subequations}
where $\zeta(s)$ is the Riemann zeta function. The increase in ``transcendentality'' of the numbers entering the PN expansion coefficients as the order increases is discussed in \cite{BiDa2.14}. Note also the half-integral contributions occuring at the leading 5.5PN order, first pointed out in \cite{Sh.al.14}. These were later shown to be closely related to gravitational-wave tails-of-tails \cite{Bl.al.14,Bl.al2.14}.

In Sec.~\ref{subsec:kappa} we used the 8.5PN-accurate formula \eqref{z1}--\eqref{coeffs} in order to approximate the surface gravity \eqref{4mu1kappa1temp} of the smaller body, to linear order in the (irreducible) mass ratio $q_\mu$, and in Eqs.~\eqref{a-d}, following the mass-ratio rescaling. Ideally, one should use numerical data for $z_{(1)}(u)$ that does not rely on a PN expansion. However, we checked that the perturbative prediction \eqref{4mu1kappa1temp} with the PN expansion \eqref{z1}--\eqref{coeffs} is fairly insensitive to the PN order at which we truncate the $\calO(q_\mu)$ contribution from $z_{(1)}(u)$. More precisely, we checked that for $q_\mu = 1/10$ the relative differences between the 6.5PN, 7.5PN and 8.5PN approximations remain below $1\%$ for $\mu\Omega \lesssim 0.16$. For larger orbital frequencies, however, the PN expansion becomes unreliable, and exact numerical gravitational self-force data would be needed.

\bibliography{}

\end{document}